\documentstyle[preprint,aps,epsfig]{revtex}
\tightenlines

\begin{document}
\draft
\title{
"Quasi Universality classes" in 2D frustrated $XY$ spin systems.
}
\author{D. Loison{\footnote
{Electronic address: Damien.Loison@physik.fu-berlin.de}}
}
\address{
{\it  Institut f\"ur Theoretische Physik, Freie Universit\"at Berlin,
Arnimallee 14, 14195 Berlin, Germany}\\
}
\maketitle

\begin{abstract}
Classical $XY$ spins on a two dimensional triangular lattice 
with antiferromagnetic interactions are reconsidered. 
We find that the
Kosterlitz-Thouless transition associated to the $U(1)$ symmetry
appears at a temperature 0.0020(2) below the Ising transition at 0.5122(1) 
associated to the $Z_2$ symmetry.
The Ising transition has critical exponents different from the
standard ones. 
Using  extensive Monte Carlo simulations
for equilibrium and dynamical properties
we show that the lack of universality observed in previous 
studies is due to finite size corrections not taken account.
Likewise the Kosterlitz-Thouless transition has 
a critical exponent $\eta\approx0.36$ larger than the corresponding 
standard value $0.25$. 
Also the helicity jump at the critical temperature is 
smaller than in the 
ferromagnetic case in disagreement with theoretical predictions.
We try using the concept of an "quasi Universality class" to
reconcile the standard critical behavior observable at higher 
temperatures with the different quasi universal one close
to the critical region.
\end{abstract}
\vspace{1.cm}
PACS number(s): 75.10.Hk, 05.70.Fh, 64.60.Cn, 75.10.-b

\newpage

\section{Introduction}

The improvement in micro-fabrication has increased greatly the 
experimental studies of the Josephson-Junction arrays of weakly 
coupled superconducting islands.\cite{Lobb88,Ling96}
Phase transitions in these arrays are similar to those in 
two dimensional $XY$ spin systems and the application of a magnetic field 
introduces additional frustration effects.
This explains partly the revival of interest in studying 
$2D$ frustrated $XY$ spin systems. 
The theoretical problem is connected to the presence of two symmetries 
and their coupling.
Indeed the continuous $XY$ symmetry leads to a Kosterlitz-Thouless transition
driven by the unbinding of vortex-antivortex pairs,
while the frustration introduce an additional Ising symmetry.
Also for Helium-3 film
an Ising symmetry due to the p-wave order parameter exists
besides the continuous symmetry of a phase like
for Helium-4 films.\cite{Halsey85,Korshunov86,Kotsubo87,Xu90}
Contrary to the three dimensional 
case, the critical region in films could be accessible to experimental 
observations.
The model we are going to analyze has also a link to the physics
of early universe,\cite{Vachaspati98} where more
complicated couplings between symmetries are considered,

Extensive research has been done on two dimensional frustrated spin systems 
like the Fully Frustrated $XY$ model 
\cite{Villain77,Teitel83,Berge86,Nicolaides91,Jose96,Thijssen90,Granato93,LeeLee94,Luo98,JLee91} 
or related Zig-Zag models,\cite{Benakly97,Boubcheur98} the triangular
model we are discussing,\cite{JLee91,Miyashita84,LeeLee98,Caprioti97} 
the $J_1-J_2$ model,\cite{Fernandez91,Loison.Simon.J1J2} the $XY$-Ising model 
\cite{Nightingale95,Granato91} and the Villain model.\cite{Olson95} 
Also other systems have the same symmetries:
the 19 vertex model,\cite{Knops94}
the $1D$ quantum spins,\cite{Granato92} 
the Coulomb 
gas representation,\cite{Yosephin85,Minnhagen85,Grest89,JRLee94}
the $XY-XY$ model.\cite{Choi85,Jeon97} or the RSOS model\cite{LeeLeeRSOS}
Therefore it is of great interest to compare the results for these 
different models in order to find out whether a universal critical 
behavior exists.

A priori only two possibilities exist for the critical behavior.
First, if the two phase transitions connected to the two symmetries 
are at the same temperature, then 
a new universal behavior should result.
Secondly if the transitions are at two different temperatures
one could expect that the transitions are of the standard Ising and 
Kosterlitz-Thouless type, especially if the transition temperatures 
are further apart. 
The presence of topological defects or vortices could  change the reasoning.
Indeed numerous studies show transitions at different critical temperatures, 
however with exponents for the Ising transition which vary from 
model to model and 
therefore cannot belong to the standard nor to one universality class.  
The reason could be the existence of a line 
of fixed points generating the various exponents found in numerical 
works.\cite{Granato91}
A second proposition assumes only one new fixed point
with the variation of critical exponents due to improper 
finite sizes corrections.\cite{Knops94}
Thirdly Olson suggested that a large screening length around the critical 
temperature of the Kosterlitz-Thouless (KT) transition prevents to see
the true standard Ising behavior.\cite{Olson95}

In this article we will show that several studies which indicate a
non-universality of the exponents are due to finite size corrections.
We are therefore led
to favor the existence of only one fixed point.
Moreover to reconcile the result of  
studies done in the finite size scaling region very close to the
critical temperature with those done at slightly higher temperatures 
we will discuss introduce the concept of an "quasi universality class"
and give a physical interpretation as function of the size of vortex.

The behavior of the Kosterlitz-Thouless (KT) transition have been less studied
numerically. Especially the critical temperature is difficult to find 
even for simpler ferromagnet $XY$ system.  
We propose a new way using the Binder parameter to overcome this difficulty 
and we obtain a critical temperature 0.002 below the Ising transition.
We obtain a helicity jump smaller than the jump of the ferromagnetic 
system and an exponent $\eta\approx0.36$ that is larger than $\eta=0.25$ for 
the ferromagnet, contrary to the theoretical 
predictions.\cite{Yosephin85,Minnhagen85,Grest89,JRLee94}

Also for the KT transition there is a 
discrepancy between simulations at or near the critical temperature
and at high temperatures. Indeed in this last region the exponent 
$\eta\approx0.22$. We introduce a "quasi universality class"
to explain this crossover.

A combination of the Metropolis algorithm and 
over-relaxation algorithm reduce the
CPU time by an order of magnitude than the metropolis alone. 
With longer simulations, up to 10 times compared to previous studies,
we gain two order better statistics.

For the determination of critical exponents we have used the 
finite size scaling method.
However, since the possible presence of a screening length
could prevent to see the "true" behavior,\cite{Olson95}
we have also utilized the properties
of the system in the short time critical dynamic 
\cite{Janssen89,Huse89,Humayun91,Zheng98,Luo97} 
which allows us to verify our results.
Both methods agree.

The outline of the article will be the following. 
Section II is devoted to the presentation
of the model.  The Ising and the Kosterlitz-Thouless 
transitions are studied in section III and IV respectively,
discussion and conclusion are disclosed in the last section.
To avoid repetition with a previous article\cite{Loison.Simon.J1J2}
we will refer often to it.

\section{model}

We study the $XY$ spins on triangular lattices with antiferromagnetic 
interactions. The Hamiltonian is given by:
\begin{eqnarray}
H&=&J\sum_{\langle ij\rangle}{\bf S}_{i}.{\bf S}_{j}\\
 &=&J\sum_{\langle ij\rangle} cos(\theta_i-\theta_j)
\end{eqnarray}
where ${\bf S}_{i}$ is a two component classical vector of length unit,
$J$ is the antiferromagnetic coupling constant ($J>0$),
$\theta$ varies between 0 and $2\pi$
and $\langle ij\rangle$ are the next nearest-neighbors.
  
The competition between the interactions gives the famous "$120^\circ$"
structure where the spins are not collinear (fig.~\ref{fig.GS}) and 
where the frustration is divided amongst all links.
Due to this structure,
the simulated lattice sizes  
must be a multiple of 3. We simulate the
sizes $L=$12, 18, 24, 36, 48, 60, 81, 105, 123, 150.
Two ground states exist, not related by a global rotation, and therefore,
in addition of the symmetry $U(1)$ from the continuous aspect of the spins, an
Ising symmetry is present. 

To compute the order parameter for the $U(1)$ symmetry we divide the lattice 
in three sublattices ($s=1,~2,~3$) with only parallel spins in the ground state.
After having calculated the magnetization of each sublattice $M_s$ we sum 
them to obtain $M$:  
\begin{eqnarray}
\label{formule.M}
M = {1 \over N} \,\sum_{s = 1}^3 |\,M_{s}| \ .
\end{eqnarray}
$N=L^2$ is the total number of the lattice sites.

The order parameter $\kappa$ of the Ising symmetry is the sum of the 
chiralities $\kappa_i$ of each cell (fig.~\ref{fig.GS}).
\begin{eqnarray}
\label{formule.kappa}
{\bf \kappa}_i &  = & \frac {2}{3 \sqrt{3}} \, \Bigl{[} {\bf S}^1_i
\times {\bf S}^2_i \, + \, {\bf S}^2_i \times {\bf S}^3_i \, + \, {\bf S}^3_i
\times {\bf S}^1_i  \Bigr{]} \ ,\\
\kappa &  = &  {1 \over N'} \,\big| \, \sum_{i} {\bf \kappa}_i \, \big| \ ,
\end{eqnarray}
where the summation is over all cells and $N'=3 N$ is their number.
The chirality $\kappa_i$ of one triangle is parallel to
the $Z$-axis and equal to $\pm1$ in the ground state only. We note that we 
could take as a definition the sum of the signs of chiralities. The result
should be similar because the length of the chiralities $\kappa_i$ is not
relevant for the critical behavior. 

To compute the critical properties we have to define for each temperature
the following quantities:
\begin{eqnarray}
\label{formuleXM}
\chi_2^M &=& {{N\langle M^2\rangle} \over {k_{B}T}} \\
\label{formuleXK}
\chi^{\kappa} &=&{ {N(\langle \kappa^{2}\rangle-\langle \kappa\rangle^2)}\over {k_{B}T}} \\
\label{formuleXK2}
\chi_2^{\kappa} &=& {{N\langle \kappa^{2}\rangle}\over {k_{B}T}} \\
\label{formuleV1}
V_{1}^\kappa&=& {{\langle \kappa E\rangle}\over {\langle \kappa\rangle}}-\langle E\rangle \\
\label{formuleV2}
V_{2}^\kappa&=&{ {\langle \kappa ^2E\rangle}\over {\langle \kappa^2\rangle}}-\langle E\rangle \\
\label{formuleV2.M}
V_{2}^M&=& {{\langle M ^2E\rangle}\over {\langle M^2\rangle}}-\langle E\rangle \\
\label{formuleUM}
U^M&=&1-{ {\langle M^{4}\rangle}\over {3\langle M^{2}\rangle^{2}}}\\
\label{formuleUK}
U^{\kappa}&=&1-{{\langle \kappa^4\rangle}\over {3\langle \kappa^2\rangle^2}}\\
\label{eqn.HELI}
\Upsilon&=&-{ \langle E\rangle \over \sqrt{3} } -{2 \over \sqrt{3}NT} 
\langle \large{[}\sum_{\langle ij\rangle} \sin(\theta_i-\theta_j)x_{ij}\large{]}^2\rangle
\end{eqnarray}
$E$ is the energy,  
$\chi$ is the magnetic susceptibility per site, 
$V_{1,2}$ are cumulants used to obtain the critical exponent $\nu$,
$U$ are the fourth order cumulants,
$x_{ij}=x_i-x_j$ where $x_i$ is the coordinate of the site $i$ following one 
axe,
$\Upsilon$ is the helicity\cite{LeeLee98} corresponding to the increment of 
the free energy for a long wavelength twist of the spin 
system,\cite{Fisher73,Ohta79}
$\langle ...\rangle$ means the thermal average.

\section{Ising symmetry}
This chapter is devoted to the Ising symmetry. The first part is related to the 
properties in equilibrium, i.e. we calculate the various quantities 
after a time $t_{th}$ to thermalize the system 
much greater than the correlation time $\tau$. The average is done 
on a time $t_{av}$ which is also much greater than $\tau$. 
In a second part we will use the short time critical dynamic 
recently introduced, i.e. the reaction of the system to a quench at 
the critical temperature from an initial state. No thermalization is used
($t_{th}=0$). 

\subsection{Equilibrium properties}
\subsubsection{Algorithm}
As explained in our previous article,\cite{Loison.Simon.J1J2} we use a 
combination of $N_{MET}$ Metropolis steps and $N_{OR}$ over-relaxation
steps.\cite{Overrelaxation}
The over-relaxation algorithm is "microcanonical" in the sense
that the energy does not change under a step. This algorithm reduces 
considerably the autocorrelation time. 
We have thus two parameters ($N_{MET}$ and $N_{OR}$) to fit in order 
to minimize
the CPU time $f_{CPU}$. 
Our implementation for the over-relaxation algorithm is six times
quicker than the Metropolis algorithm:
\begin{eqnarray}
f_{CPU}(N_{MET},N_{OR})&=& \tau\,(N_{MET}+{N_{OR}\over 6})\\
\label{eqn.f.CPU}
&=& \tau\,N_{MET}\,(1+{N_{OR} \over 6\, N_{MET}}) 
\end{eqnarray}

In Fig.~\ref{fig.NMET.NOR} we have plotted the autocorrelation time $\tau$
for the chirality $\kappa$, at the critical temperature $T=0.5122$ calculated
below, multiplied by $N_{MET}$, as function of $N_{OR}/N_{MET}$ in a
log-log plot for a lattice size $36$. 
$\tau$ is calculated with the method explained in the Appendix of our 
previous article.
The data are well described by:
\begin{eqnarray}
\label{eqn.tau.MET}
\tau \, N_{MET}=a_L \, \left( {N_{OR} \over N_{MET}} \right)^{b_L}
\end{eqnarray}
with $a_{36}=160$ and $b_{36}=-0.57$. 

Using (\ref{eqn.tau.MET}) it is not difficult to show 
 that the minimum of $f_{CPU}$ (\ref{eqn.f.CPU}) occurs  for:
\begin{eqnarray}
{N_{OR} \over N_{MET}}={-6\,b \over 1+b} \, ,
\end{eqnarray}
which is 7.95 for $L=36$. 
We have determined $b$ for several lattice sizes $L$.
If we divide the result by $L$, 
we obtain the ratio $c_L$: $c_{12}\sim0.18$,
$c_{24}\sim0.28$, $c_{36}\sim0.24$, $c_{48}\sim0.21$, $c_{60}\sim0.21$.
This ratio is nearly constant whatever ${N_{OR} \over N_{MET}}$ and $L$ are. 
This is in accordance with a conjecture of Adler\cite{Adler81} stating 
that this ratio is 
proportional to the correlation length, i.e. the size in the finite 
size region where we have done the simulations.
We have chosen $N_{MET}=1$ and
$N_{OR}\approx 2 L/9$ for the simulations.

We show in Fig.~\ref{fig.tau.L} $f_{CPU}$ for the chirality  
as function of the size of the lattice $L$ in a log-log plot for various
algorithms at the critical temperature $T=0.5122$. A similar behavior is 
obtained for $M^2$, i.e. the order parameter associated to the $U(1)$ 
symmetry.
$A_{0}$ is the metropolis algorithm alone, 
$A_{1}$ is used in combination with one step of an over-relaxation algorithm, 
$A_{cL}$ in combination with $0.22 L$
steps of an over-relaxation algorithm, while $A_W$ correspond to the use 
of the Wolff cluster algorithm.\cite{Wolff89}

The slope of $A_{0}$ gives us the dynamical exponent 
\begin{eqnarray}
z^\kappa=2.30(4).
\end{eqnarray}
We will discuss the value of this exponent at the end of this
section. If we compare $f_{CPU}$ for $A_{0}$ and
$A_{cL}$ we observe that the gain is about 10,
which means that for the same time of 
simulation we obtain ten times better statistics (i.e. "independent" data).
We note that Wolff's algorithm is less effective than the Metropolis algorithm.
This is understandable because this algorithm uses only one link at each time
to construct the cluster and we know that three links must be taken into 
account (at least one cell), therefore the algorithm can not generate
the "good" cluster. Indeed each cluster has about $80\%$ of the sites
of the lattice, which is too high.
Moreover even if we would be able to construct a good cluster,
there is no guarantee that the method to flip the spins inside would be
efficient.\cite{Sokal91}

\subsubsection{Errors and details of the simulation}
We follow the procedure explained in the Appendix of our previous 
article.\cite{Loison.Simon.J1J2}

We use in this work the histogram method\cite{Ferren88} 
which allows us from a simulation done at $T_{0}$
to obtain thermodynamic quantities at $T$ close to $T_{0}$.
However to reduce the systematic
errors we do not save histograms for $\langle M\rangle$ , $\langle M^2\rangle$ \dots as function 
of energy but save the data, that is the energy $E$, the magnetization $M$ and
the chirality $\kappa$. To avoid the use of a large space on the hard disk 
the data is saved only every $\tau_s$ sweeps (see table~\ref{table.tau}). 
This method slightly increases our statistical
errors (approximative of 20\%). 
However the systematic errors decrease and we have a better
control of the total errors.

Since the data of a Monte-Carlo simulation are not independent,
the calculation of errors must be done carefully. For simple quantities
like the magnetization, the calculation is done with the standard formula of
statistic but with the number of "independent" data $t_{ind}$ equal to 
${N_{MC}/\tau_s \over 2\tau/\tau_s+1}={N_{MC}\over 
2\tau+\tau_s}$.\cite{Binder73} $N_{MC}$ is the number of Monte Carlo steps. 
However for more complicated quantities we have to 
consider the correlations between the components of the formula and use,
for example, the  jackknife procedure.\cite{Jackknife}
Formula (A8-A14)  of our previous article\cite{Loison.Simon.J1J2} 
can be used but we have to change $N_{MC}$ to $N_{MC}/\tau_s$ 
and $\tau$ to $\tau/\tau_s$.
However the error for the helicity is not given.
The helicity can be written:
\begin{eqnarray}
\Upsilon=\langle A\rangle - \langle B\rangle
\end{eqnarray}
where $A$ and $B$ are given by (\ref{eqn.HELI}).
Applying the same method as in Refs.\cite{Loison.Simon.J1J2}, it is not 
difficult to obtain:
\begin{eqnarray}
\Delta\Upsilon^2={2\tau+\tau_s \over N_{MC}}
\large{[}\langle A^2\rangle -\langle A\rangle^2 + \langle B^2\rangle - \langle B\rangle^2 - (\langle AB\rangle-\langle A\rangle\langle B\rangle) \large{]} \  . 
\end{eqnarray}

The simulations have been done using sizes between 12 to 150.
In the table~\ref{table.tau} we gave some details of the simulations
where $t_{th}$ is the number of Monte Carlo steps (i.e. one Metropolis step
followed by $2L/9$ over-relaxation steps) to thermalize the system;
$t_{av}$ is the number of Monte Carlo steps to average. The third 
column gives the autocorellation time, the fourth the time between
two consecutive measures $\tau_s$. The last column is the number of 
"independent" data. We report the last line in 
table~\ref{table.revue.I}, it should be compared to previous studies. 
Our statistic is two orders greater than previous studies for 
similar sizes. This allows us to obtain better precisions for the
quantities, typically one or two orders smaller, and therefore 
the critical exponents are more reliable.
In particular we will see that the finite size corrections are important 
and this explains the variation of exponents found in previous studies 
(see discussion at the end of this section).

\subsubsection{Results}
Our first task is to find the critical temperature $T^\kappa_c$.
The most effective way is to use Binder's cumulant (\ref{formuleUK})
in the finite size scaling (FSS) region.
We record the variation of $U^\kappa$
with $T$ for various system sizes in Fig.~\ref{figCM}
and then locate $T^\kappa_c$ at the intersection
of these curves\cite{BinderU} since
the ratio of $U^\kappa$ for two different lattice sizes $L$
and $L'=bL$ should be 1 at $T^\kappa_c$.
Due to the presence of residual corrections to finite size scaling,
one has actually to extrapolate the results taking the limit
(ln$b$)$^{-1} \rightarrow  0$ in the upper part of Fig.~\ref{figTc2.all}. 
We observe a strong correction for the small sizes. However for the biggest
sizes the fit seems good enough and we can
extrapolate $T^\kappa_c$ as
\begin{equation}
\label{eqn.Tc.K}
T_c^\kappa= 0.5122(1)\ .
\end{equation}
We note from the figure that it is a lower bound. 
The estimate for the universal quantity $U^\kappa_*$ at the
critical temperature is
\begin{equation}
U^\kappa_*=0.632(2).
\end{equation}
This value is far away from the two dimensional Ising value 
$U_*^{Ising}\sim0.611$\cite{Kamierenarz-Blote} 
which is a strong indication that the Universality
class associated to  the chirality order  parameter is not of Ising type. 
We note that it is not compatible with the value of the $J_1-J_2$
model 0.6269(7)\cite{Loison.Simon.J1J2} which, in the standard formulation,
means that the two systems belong to two different Universality classes.
However since this transition is a coupling between two it is not certain
that this quantity stays universal.

At $T=T^\kappa_c$ the critical exponents can be determined by log--log fits. 
We obtain $\nu^\kappa$ from $V_1^\kappa$ and $V_2^\kappa$ (Fig.~\ref{figV}),
$\gamma^\kappa/\nu^\kappa$
from $\chi^\kappa$ and $\chi_2^\kappa$ (Fig.~\ref{figX.all}),
and $\beta^\kappa/\nu^\kappa$ from $\kappa$ (see Fig.~\ref{figM}).
We observe in these figures a strong correction to a 
direct power law. It is worth noticing  however that $X_2^\kappa$ shows 
smaller corrections.
Using only the three (four for $\chi_2^\kappa$) 
largest terms ($L=$105, 123, 150)
we obtain: 
\begin{eqnarray}
\label{formule.nu.K}
\nu^\kappa&=&0.815(20)\\
\label{formule.gammasnu.K}
\gamma^\kappa/\nu^\kappa &=& 1.773(9) \\
\label{formule.betasnu.K}
\beta^\kappa/\nu^\kappa &=& 0.110(6)\ .
\end{eqnarray}
The uncertainty of $T^\kappa_c$ is included in the estimation of the errors.
The large values of our errors are due to the use of only few sizes 
for our fits.
The values obtained for four consecutive sizes are interesting. For
the exponent $\nu$ we obtain: 
$\nu_{12-18-24-36}=0.909$, $\nu_{18-24-36-48}=0.909$,
$\nu_{24-36-48-60}=0.903$, $\nu_{36-48-60-81}=0.884$,
$\nu_{48-60-81-105}=0.862$,
$\nu_{60-81-105-123}=0.835$, $\nu_{81-105-123-150}=0.822$.
These values cover a large range of the data obtained by previous
studies (see table~\ref{table.revue.I}) 
and we strongly suspect that the lack of Universality (at least in 
the critical exponents) is related to the corrections 
previously ignored.
In the conclusion of this section we will show which important physical
informations we are able to obtain from the sign of these corrections.
 
We have tried to introduce a correction to calculate the
exponents, for example for 
$V_1^\kappa=(1+L^{-\omega^\kappa})L^{1/\nu^\kappa}$,
we  obtain
$\omega^\kappa=1.2(5)$ and values for critical exponents fully compatible
with (\ref{formule.nu.K}-\ref{formule.betasnu.K}). 

The values given in (\ref{formule.nu.K}-\ref{formule.betasnu.K}) use the
properties of the free energy at the critical temperature. But an error
on $T^\kappa_c$ leading to an error on the exponents,  
it is therefore advisable to
find them without the help of $T^\kappa_c$. This can be done using 
the whole finite size scaling region with the method given 
in.\cite{Loison.O6.Ferro} 
It consists to plot, for example, the susceptibility 
$\chi^\kappa L^{-\gamma^\kappa/\nu^\kappa}$ as function of $U^\kappa$, choosing
the exponents so that the curves collapse. 
This fit is more reliable than the fit at the
critical temperature since as it does not depend only on the results at
$T^\kappa_c$ but on a large region of temperature. 
However the errors are a little bit larger. We show in 
Fig.~\ref{X2.1}-\ref{X2.3} the results for three choices of 
$\gamma^\kappa/\nu^\kappa$ 1.82, 1.78 and 1.74. 
Obviously the second is the best and we obtain  
$\gamma^\kappa/\nu^\kappa=1.78(2)$ 
compatible with the result of (\ref{formule.gammasnu.K}). 

In Fig.~\ref{fig.eta} we have plotted the exponent 
$\eta^\kappa=2-\gamma^\kappa/\nu^\kappa$ 
calculated with a direct fit of the susceptibility 
(like in Fig.~\ref{figX.all}) but changing the temperature. In the same
diagram we have plotted the result ($\Delta\eta^\kappa$) 
found by the fit of $\chi^\kappa$ as function of $U^\kappa$. This gives an 
estimate of the critical temperature (see figure) compatible with $T_c^\kappa$.

The results are summarized in table~\ref{table3}.
In the next section we will try to compute the properties of this
transition using the short time critical behavior. This method does not 
suffer from the same problems as the finite size scaling method and allows us
to check our results. Finally we will compare our results with those 
of previous studies.

\subsection{Short time critical behavior}
In this section we want to study our system using the dynamical behavior
near the critical temperature. This particular method consists of
quenching the system from zero temperature to the critical temperature and
studying the short time dynamics,
i.e. before reaching the equilibrium properties.
It seems strange that the system shows critical behavior 
although it has not reached equilibrium. Indeed for a long time 
this fact has not been utilized in numerical simulations.
However it has been demonstrated that
between a time $t_i$ and a longer time $t_f$ a region exists where the
calculation of the critical properties of the transition is possible 
\cite{Janssen89,Huse89,Humayun91,Zheng98,Luo97}
(see the review of Zheng\cite{Zheng98}). In this region 
the system has lost the non-universal informations ($t>t_i$) but the 
correlation length $\xi(t)$ is much smaller than the correlation length
$\xi(\infty)$ of the system in the equilibrium phase.
Since we are at $T_c$ the correlation length
$\xi(\infty)$
is infinite but in our simulation it is upper-bounded by the size $L$ 
of the lattice and the condition written as $\xi(t) \ll L$. 
After $t_f$ this condition is not satisfied and the system shows a crossover
to equilibrium properties.
In the region $t_i < t < t_f$ the system shows very simple power law 
even for the Kosterlitz-Thouless transition.\cite{Zheng98,Luo97}

The quantities to compute are the same
as in formula (\ref{formule.M}-\ref{formuleUK}) but the average $\langle \dots\rangle$
is now taken on different realizations, i.e. starting from the ground
state with different random numbers for the Metropolis procedure. 

To determine the range of time and the size $L$ that we use in the 
simulations we have to find $t_f$ where the system begins to show a 
crossover to the equilibrium properties. In Fig.~\ref{fig.a4} we
show $\langle \kappa(t)\rangle$ as function of the time $t$ for different sizes $L$
at the critical temperature $T=T_c^\kappa=0.5122$ in a log-log plot. 
We observe that for small systems $t_f\approx100$. It grows for larger
systems 
and since the sizes $L=201$ and $L=300$ give similar results for 
$\langle \kappa(t)\rangle$ in Fig.~\ref{fig.a4} $t_f$ is bigger than $10^4$.
Therefore we can use the entire region 
to calculate the critical exponents for the greatest size.

We have studied systems with size $L=300$ for a simulation time $t$
up to 10,000 and we have averaged over 6000 realizations.
This reflects more than one order statistics better than previous studies
on the fully frustrated $XY$ model (FF$XY$).\cite{Luo98}
For the algorithm we use only 
the Metropolis algorithm (Model A in the classification of Ref. \cite{Hohenberg77}).
Here we do not use the over-relaxation algorithm.
The errors are calculated dividing the data in three sets of 2000.

We want now to verify the critical temperature $T_c^\kappa=0.5122$ found
in the previous section. For this purpose we have simulated our system for
three different temperatures at and around this value. We know that at 
the critical temperature the magnetization shows a power law behavior,
 a straight line in a log-log plot. For $T\ne T_c$ the system shows
important corrections as observed in Fig.~\ref{fig.a2}.

With our knowledge of the critical temperature found above, we are 
able to calculate very precise exponents.
In the short time critical dynamic the quantities 
(\ref{formule.M}-\ref{formuleUK}) have at $T_c$ the behavior:
\begin{eqnarray}
\label{form.M.t}
\langle \kappa(t)\rangle &\propto& t^{-\beta^\kappa/(\nu^\kappa z^\kappa)} \\ 
\label{form.X.t}
\chi^\kappa &\propto& t^{\gamma^\kappa/(\nu^\kappa z^\kappa)} \\ 
\label{form.V1.t}
V_1^\kappa &\propto& t^{1/(\nu^\kappa z^\kappa)} \\ 
\label{form.U.t}
U^\kappa &\propto& t^{d/z^\kappa} 
\end{eqnarray}
where $z^\kappa$ is the dynamical exponent and $d$ the dimension of the space, 
i.e. $d=2$.

We first compute  $z$ using (\ref{form.U.t}).
In Fig.~\ref{fig.U.t} we have plotted $U^\kappa$ as function of $t$.
We see that from $t_i\sim100$ the curve shows a linear behavior. Using a fit 
from $t=300$ to $t=10,000$ (shown by the two arrows) we are able to obtain
\begin{eqnarray}
\label{form.z}
z^\kappa=2.39(5).
\end{eqnarray}
This value is consistent with the value found in equilibrium properties
($z_{eq}^\kappa=2.30(4)$), but we are more confident in the result of 
dynamical properties which are less subject to systematic errors (see
Appendix of Ref. \cite{Loison.Simon.J1J2}). It is also consistent with the value 
of $z=2.31-2.36$ found studying the equilibrium properties
of the Fully Frustrated $XY$ model.\cite{Pawig98}
Nevertheless it is in contradiction
with the value $z^\kappa=2.17(4)$ found studying the dynamical properties 
of the last model.\cite{Luo98}
However a slight change of the critical temperature could have a
strong influence on $z$ and the calculation  
have been done with more than one order 
less statistics than our work.  

To obtain the other exponents we could use the formula 
(\ref{form.M.t}-\ref{form.V1.t}), however we prefer to obtain directly 
$\nu^\kappa$ and $\gamma^\kappa$ plotting $U^\kappa$ and $\chi^\kappa$
as function of $V_1^\kappa$ and $\eta^\kappa=2-\gamma^\kappa/\nu^\kappa$
plotting $\chi^\kappa$ as function of $U^\kappa$.
This has been done in Fig.~\ref{fig.U.X.V1} and Fig.~\ref{fig.X.U}.
Fits have been done for $300<t<10,000$ represented by the arrows. 
We have also calculated $\beta^\kappa$ plotting $\langle \kappa\rangle$ as function 
of $V_1^\kappa$ (not shown). We obtain:
\begin{eqnarray}
\nu^\kappa&=&0.818(9)\\
\gamma^\kappa&=&1.445(20)\\
\beta^\kappa&=&0.0967(13)\\
\eta^\kappa&=&0.235(7)
\end{eqnarray}
Our results are summarized in table~\ref{table3}.
The exponents calculated by the two methods
(equilibrium and dynamical) agree very well.
However we would prefer the results from short time critical dynamics
since they are less sensitive to the finite sizes corrections.

\subsection{Comparison with previous studies}
Before discussing the interpretation of our results, we will compare 
them to those found previously on various models.

In table~\ref{table.revue.I} we review the numerical studies
related to our work.
We think that the errors quoted in this table
are mostly too optimistic estimates.

We first compare our work with previous simulations using the same methods
as this work (Monte Carlo Finite Size Scaling-MC FSS).
There are two important columns:
the maximum size $L_{max}$ used (fourth column) and the number of independent
data $t_{ind}$ (seventh column). Indeed we have seen that 
strong finite size corrections exist in our system  and we cannot see them if 
the largest size is too small. When comparing the calculations with a similar
largest size ($L\sim150$),\cite{Boubcheur98,Caprioti97,Fernandez91} 
we observe that our statistics ($t_{ind}$) is between 100 to 1000 times
better. This allows us to see the corrections not seen before and to explain the
lack of Universality observed in previous studies (see subsection "equilibrium
properties"). The results, for the triangular lattice, in favor of the Ising
ferromagnetic Universality class ($\nu=1$)\cite{Caprioti97} are not reliable 
due to too small statistics ($t_{ind}=15$).  

For the other methods none of the authors have reported the strong corrections
which should exist.  
Nevertheless almost all results are compatible with ours ($\nu\sim 0.795-0.82$)
and are in favor of a single Universality class. Only results on
the $XY$-Ising model seem to show different Universality classes but
simulation sizes are so small that results are not conclusive.

Our results are thus strongly in favor of the picture of\cite{Knops94}
which suggests the phase diagram drawn in Fig.~\ref{fig.I-XY}. 
$A$ and $C$ are free  parameters. Similar phase diagrams appear for
the $J_1-J_2$ model,\cite{Loison.Simon.J1J2}
the Zig-Zag model\cite{Benakly97,Boubcheur98}
and the RSOS model\cite{LeeLeeRSOS}.
A variation in free parameters changes only the initial point of the 
renormalization group flow on the line $PT$ but all trajectories 
converge to the same fixed point $F$, i.e. a single Universality class. 
The finite size 
corrections will be more or less important following the initial point.
To verify this interpretation it would be useful to test several 
initial points of the line $(PT)$ 
of the $XY$-Ising model with similar sizes as in this work 
($L_{max}\sim150$) to see if the systems belong 
to the same Universality class and,
maybe more interesting, to observe the corrections to scaling. 
The picture should be very similar to the Potts model with disorder where
this crossover, and therefore the corrections, have been 
observed.\cite{Picco98} 
One other numerical study also interesting should be the calculation of
the properties
of the 19-vertex model by the Density-Matrix Renormalization Group (DMRG).
The model has already been studied by Monte Carlo Transfer Matrix 
\cite{Knops94} but only for small sizes ($15^2$). On the contrary 
the DMRG allows to treat $L$ chains of infinite sizes. 
We note that the method has been proved to be very efficient for the 
19-vertex for the ferromagnetic system (i.e. with different internal 
parameter).\cite{Honda97}

We now discuss the results of the Monte Carlo (MC) simulation 
in the high temperature (HT) phase. In this region we have to keep
the correlation length $\xi$ much smaller than the  size $L$ of the 
lattice: $\xi\ll L$.
studies on the FF$XY$ model have been done by Nicolaides\cite{Nicolaides91}
and by Jose and Ramirez.\cite{Jose96} They found two very different
results. However with a close look on their simulations we have  found 
that  the results of Nicolaides were not reliable and his errors more than
strongly underestimated. Therefore the HT MC seems 
in favor of a new Universality class but with an exponent $\nu\sim0.89$
larger than found in the finite size scaling MC ($\nu\sim0.81$). 
On the other hand Olson\cite{Olson95} has done a HT MC on the Villain
model, i.e.
where the spinwave are decoupled from the vortex, and has observed
a very interesting behavior. He proposed that  a 
screening length $\lambda=\xi_{KT}$ exists, 
due to the Kosterlitz-Thouless transition and,
that in addition to the condition $\xi_{Ising} \ll L$,  the sizes 
have to satisfy the condition $\xi_{KT}\ll L$.
In this case the system shows a standard Ising behavior $\nu=1$ while
it is in the crossover to this behavior if the condition $\lambda\ll L$
is not respected. Indeed he found some indication that the exponents
$\nu\sim1$ when the two conditions are satisfied. However since the numerical
results are not extremely precise, further simulations are needed to
prove this interpretation. 

Let's assume that this last interpretation is correct.
It does not prove that there is no new fixed points contrary to the 
claim of Olson.  Indeed we have shown in this work and in 
\cite{Loison.Simon.J1J2} that the corrections lower the result for
$\nu$ (from 0.90 to 0.80) and therefore that our system is not in the 
crossover to the ferromagnetic Ising behavior $\nu=1$.  
We then conclude that a new "stable" fixed point exists 
for a range of size $L$. We call this new fixed 
point an "quasi fixed point" by similarity with the 3D case where
an "almost second order" exists for a certain range of size $L$ before 
showing the "true" first order transition.\cite{Zumbach93,LoisonSchotteHei}
This can be understood if we admit that the Kosterlitz-Thouless transition 
coupled the spins at large distance by the intermediate of $\xi_{KT}$. 
Therefore it has tendency to bring the
critical behavior of the Ising symmetry to the mean-field solution 
($\nu=0.5$) exact for infinite interactions.  
Since $\xi_{KT}$ is finite above the critical temperature the 
thermodynamic limit of this "quasi fixed point" is not stable.

To be exhaustive we mention the study of the transition 
$XY$-Potts($q$).\cite{Loison.XY.Potts} We have shown that the transition is of
first order whenever $q \ge 3$. In this case the transition will stay of first
order even in the thermodynamic limit, since the correlation length is finite.
Therefore the new fixed point (or his absence) is stable. This fact is rather
against the interpretation of Olson but it is possible 
that the stability of the new fixed point changes between
$q=2$ (Ising) and $q=3$.
On the contrary it would say that for the $XY$-Ising model, 
the two behaviors in the finite size scaling region ($\nu\approx0.80$) 
and far away of the critical temperature ($\nu=1$) are different and
stable in the thermodynamic limit.

\section{$U(1)$ symmetry}

We will present in this section our results for the 
Kosterlitz-Thouless (KT) transition\cite{Thouless73,Kosterlitz74}
associated to the $U(1)$ symmetry.
This transition is characterized by 
the unbounding of the vortex-antivortex pair at the critical temperature.
At low temperature there are only some pairs of vortex-antivortex
while at high temperature alone vortex (or antivortex) can exist.
This picture is considered as correct in the ferromagnetic and frustrated
case.
However the two cases could show some differences for the 
jump of the helicity at the critical temperature $T_c^M$ and for the 
exponent $\eta$. 

The helicity $\Upsilon$ is the answer of the system to a twist in one direction.
It has been shown that at $T_c^M$ this quantity jump from 0 to an 
universal quantity\cite{Kosterlitz74} for all ferromagnetic systems. 
Moreover the behavior of $\Upsilon$ as function of the lattice size
has been predicted.\cite{Weber88} The use of $\Upsilon$ has been proved
to be very useful for the ferromagnetic case and in particular to determine
the critical temperature.\cite{Olson95.O2.Ferro}
The situation is not so clear for frustrated systems. It has been suggested 
that the jump at $T_c^M$  could be "non universal" i.e. different from the
ferromagnetic case\cite{Yosephin85,Minnhagen85,Grest89,JRLee94,Choi85,Jeon97}
and no scaling with the system size has been proved. 
Therefore rather than using the helicity to obtain informations
on the transition, we will use a new method for the KT transition 
that we have introduced in.\cite{Loison.Simon.J1J2,LoisonO2Ferro} 
It consists in using Binder's cumulant
to study this transition. It was proved in these articles that, contrary to
the common belief, the Binder cumulant for
ferromagnetic $XY$ systems
crosses for different sizes, allowing thus  an estimate of the
critical temperature and moreover an estimate 
of the exponent $\eta$ without the precise
knowledge of the critical temperature.
The ferromagnetic $\eta$ has been proved equal to $1/4$
\cite{Kosterlitz74} with logarithm corrections\cite{Amit80} 
while  it has been predicted to be smaller in frustrated 
cases.\cite{Yosephin85,Minnhagen85,Grest89,JRLee94,Choi85,Jeon97}

In this section we will first present our results for the equilibrium 
properties and second verify them using the short time dynamical 
properties.

\subsection{Equilibrium properties}

\subsubsection{Critical temperature and critical exponent $\eta^M$}
We proceed in a similar way as for the Ising order parameter, replacing
$\kappa$ by $M$.
We record the variation of $U^M$ (\ref{formuleUM}) with the temperature
for various system sizes in Fig.~\ref{figCM.KT}. 
We want to underline the differences between our result in the frustrated case
and in the ferromagnetic case: 
the crossing region is one order of magnitude less than for
the standard $XY$ model 
(compare with Fig.~1 of Ref. \cite{LoisonO2Ferro}). 
We then locate the intersection of these curves and plot
the results in the lower part of Fig.~\ref{figTc2.all}. 

Let us  first  consider a power law behavior at $T>Tc$ for this system. 
We then have 
to consider a linear fit for (ln$b$)$^{-1} \rightarrow 0$. We observe 
corrections for the smallest sizes $L=$12, 18 and 24 
 but the others seem to converge
to the temperature 
\begin{equation}
\label{eqn.Tc.M}
T_{c}^M= 0.5102(1)\ .
\end{equation}
We note from the figure that it is an upper bound for the critical
temperature, i.e. it cannot reach $T_c^\kappa=0.5122(1)$ and therefore we 
obtain two transitions.

Secondly we consider  the behavior to be exponential as in the standard 
$XY$ model. In this case 
Fig.~2 of Ref. \cite{LoisonO2Ferro} shows that a linear fit could be wrong
and that a "crossover" to a different critical temperature could be 
observed for bigger $b$, i.e. greater sizes. However contrary to 
the ferromagnetic $XY$ model, the region of crossing is very small and the
different linear fits tend only to one critical temperature. 
We observe the same behavior for the $J_1-J_2$ model.\cite{Loison.Simon.J1J2}
We conclude that the linear fit works well enough
and we will show below
strong arguments in favor of the temperature (\ref{eqn.Tc.M}).

With the help of the critical temperature we have found an estimate 
of $U^M$ at $T_c^M$ fitting the value with a law $U^M=U^M_*+aL^{-\theta}$:
\begin{eqnarray}
U^M_*=0.6497(12)\ .
\end{eqnarray}

The exponent $\eta$ is obtained by a log-log fit of 
$\chi_2^M$ as function of $L$ shown in Fig.~\ref{figX.all}.
We obtain 
\begin{eqnarray}
\label{formule.gammasnu.M}
2-\eta^M&=&1.635(10)\\
\label{formule.eta.M}
\eta^M&=&0.365(10)
\end{eqnarray}
The fit has been done using only the four largest sizes 
($L=$ 81, 105, 123 and 150), disregarding the smallest sizes which show
small corrections.  We note that for small sizes $\eta^M$ is smaller,
for example if we use only the sizes from 12 to 60 we obtain $\eta^M=0.33$.
Therefore if we take into account the corrections, the exponent moves away
from the standard value 0.25 for the ferromagnetic systems. This shows
that we are not in a crossover to this last behavior.
Our value is in agreement with those found for the 
$J_1-J_2$ model.\cite{Loison.Simon.J1J2}
Our results are summarized in table~\ref{table3}.

From a theoretical point of view the $KT$ transition has  an 
exponential behavior, i.e.  a correlation length of the form 
$\xi\sim\exp[B_0\,(T-T_c)^{-\nu}]$,
however a power law behavior like
($\xi\sim(T-T_c)^{-\nu}$) can not be excluded {\it numerically}. In the
latter case  
the critical exponent $\nu$ can be 
calculated with the cumulant $V_2^M$ (\ref{formuleV2.M}). We have obtained
$\nu^M=1.18(10)$ and the results show important corrections. This value being
in contradiction with the value $\nu=0.92(3)$ found in the $J_1-J_2$ model, 
we can exclude a power law behavior.

As for the Ising order parameter, the calculation of the exponents has been
done at the critical temperature but an error on $T_c^M$ leads to errors 
on the exponents, it is then interesting to
find them without the help of $T^M_c$. This can be done using the same method 
as described before.
We have shown in Ref. \cite{LoisonO2Ferro} that this method is accurate enough
in order to obtain $\eta$ whatever the type of the behavior is (power law
or exponential). In Fig.~\ref{X2.1.KT}-\ref{X2.3.KT} we show our results
for three values of $\eta^M$, 0.34, 0.375, 0.41.
Obviously the second value is the best and obtain:
\begin{eqnarray}
\label{formule.eta2.M}
\eta^M=0.375(20) 
\end{eqnarray}
which is compatible with (\ref{formule.eta.M}).

In Fig.~\ref{fig.eta} we have plotted the exponent
$\eta^M$
calculated with a direct fit of the susceptibility
(like in Fig.~\ref{figX.all}) but changing the temperature. In the same
diagram we have plotted the result ($\Delta\eta^M$)
found by the fit of $\chi_2^M$ as function of $U^M$. This gives an
estimate of the critical temperature (see figure) compatible with $T_c^M$.

We now want to demonstrate that the two transitions cannot appear at the same 
temperature. First we have shown in Fig.~\ref{figTc2.all} that 
$T_c^\kappa=0.5122(1)$ is a lower bound for the critical temperature 
while $T_c^M=0.5102(1)$ is a upper bound. Moreover
if the Ising transition appears at $T_c^M=0.5102$, the exponent
$\eta^\kappa$ must be equal to 0.115 ($\gamma^\kappa=1.885$) 
by a direct log-log fit at this temperature (see Fig.~\ref{fig.eta}).
In Fig.~\ref{X2.4} we have plotted $\chi^\kappa$ as function of $U^\kappa$
for this value. The curves do not collapse.  
Now if the Kosterlitz-Thouless transition appears at $T_c^\kappa=0.5122$ 
then $\eta^M=0.461$ (see Fig.~\ref{fig.eta}) and the curves $\chi_2^M$ as 
function of $U^M$ do not collapse in Fig.~\ref{X2.5.KT}.
Therefore, even if the critical temperatures are different by only 0.4\%,
we are able to conclude that the KT transition appears at lower temperature 
than the Ising transition. 
Moreover we want to stress the fact that the phase diagram of the
FF$XY$\cite{Berge86}
is not in contradiction with our picture. In this work the authors have
shown that with varying a parameter ($J'/J$)
it is possible to decouple the two symmetries with the Ising transition
appearing at lower temperature than the KT transition.
However in this case the Ising transition is due to the contraction to 0
(or $\pi$) of the turn angle between the spins and not to the mixed of
different chirality signs.\cite{Loison.Q99} Therefore two physical
properties exist near $J'/J=1$ which
lead to a complicated phase diagram with a cross of the critical lines.

\subsubsection{Helicity}
We present know our results for the helicity $\Upsilon$ (\ref{eqn.HELI}). 
In Fig.~\ref{fig.HELI} we have plotted  $\Upsilon$ as function of the 
temperature. $T_s$ is the temperature of simulation, $T_c^M$ is our estimate
for the critical temperature from above. $2T/\pi$ is the universal jump 
for a ferromagnetic system. 

$T_0$ is the best fit of $\Upsilon$ with the scaling form as function of 
the system size $L$ valid in ferromagnetic systems\cite{Weber88}:
\begin{eqnarray}
\label{eqn.HELI.L}
\Upsilon(L)=\Upsilon_{jump}(1+{1 \over 2\ln L+c})
\end{eqnarray}
where $\Upsilon_{jump}$ is the jump at $T_c$ and $c$ a free parameter.
We note that our results are very similar to 
those of Lee and Lee.\cite{LeeLee98} Since it is difficult to obtain,
with the histogram method, informations for the biggest sizes too far away 
from the temperature of  simulation ($T_s=0.511$)
we take their result for $T_0$:
\begin{eqnarray}
T_0=0.501(1) \ .
\end{eqnarray}
This value does not agree with $T_c^M=0.5102(1)$ found above.
If we admit $T_0$ as critical temperature the exponent $\eta^M$ must be 
equal to 0.22 (see Fig.~\ref{fig.eta}) and the curves $\chi_2^M$ as
function of $U^M$ do not collapse in Fig.~\ref{X2.4.KT}. 
Therefore we can doubt the validity of (\ref{eqn.HELI.L}) for the 
frustrated case.

More interesting is the comparison of $\Upsilon$ at $T_c^M=0.5102$ with
the ferromagnetic jump $\Upsilon_{jump}^{ferro}=2T/\pi$.
In Fig.~\ref{fig.HELI.Tc} we have plotted these two quantities as functions of 
the size $L$.
Our results suggest a jump at the critical temperature smaller than 
the ferromagnetic jump in contradiction of the suggestion 
of.\cite{Yosephin85,Minnhagen85,Grest89,JRLee94,Choi85,Jeon97}
This surprising result could induce some doubts about our previous results.
In order to verify them, we will present now the dynamical properties of the KT 
transition.

\subsection{Short time critical behavior}
This analysis is very similar to those done for the Ising symmetry.
Zheng and co-workers\cite{Zheng98,Luo97} have proved that a system quenched from
a zero temperature to the critical temperature $T_c$ shows
that a KT transition has a similar picture as the Ising transition, i.e. 
a power law behavior as function of the time $t$:
\begin{eqnarray}
\label{form.M.t.KT}
\langle M(t)\rangle &\propto& t^{-\eta^M/(2 z^M)} \\
\label{form.U.t.KT}
U^M &\propto& t^{d/z^M}
\end{eqnarray}
If the system is not quenched to the critical temperature corrections
are present and  the behavior ceases to be linear in a log-log plot.

In Fig.~\ref{fig.a3} we have plotted $\langle M\rangle$ as function of $t$ for different
temperatures. We observe a linear behavior for $T_c^M=0.5102$ and 
clearly  a deviation from this behavior for the other temperatures 
and particular for $T_0=0.5010$.  
This is a strong indication that $T_c^M$ is the good choice of the 
critical temperature.

We have calculated the exponents $\eta^M$ and $z^M$ at $T=T_c^M$ from
(\ref{form.M.t.KT}-\ref{form.U.t.KT}) by 
similar methods used for the Ising symmetry. Since the average has been
done only on one sample of 500 configurations we are not able to calculate
the errors. We obtain:
\begin{eqnarray}
z^M&=&2.10 \\
\eta^M&=&0.36
\end{eqnarray}
The value of $\eta^M$ is in agreement with those found in equilibrium 
(see table~\ref{table3}).
 
In conclusion we found that, numerically, the Kosterlitz-Thouless
in the Finite Size Scaling (FSS) region has a complete different behavior
than for ferromagnetic systems.
We found an exponent $\eta^M\sim0.36$ larger and a jump of
the helicity smaller in frustrated case.

Our exponent $\eta^M$ is in agreement with our results for the $J_1-J_2$
model $\eta^M_{J_1-J_2}\sim0.35$ (see table~\ref{table3}) with a similar
method as this work, and with the result found in the FF$XY$,
$\eta_{FFXY\sim0.34}$, using a transfer matrix 
method.\cite{Knops94}
 However they are in  disagreement with the values found
in the high temperature (HT) region $\eta^M=0.20-0.25$.\cite{Jose96,Olson95}
Although new and surprising, we believe that our results are reliable 
and present hereafter the numerical arguments in favor of them:
\begin{itemize}
\item[1.] Fig.~\ref{figTc2.all} is very suggestive for the value of
the critical temperature and shows that the behavior of the Kosterlitz-Thouless
transition in frustrated systems is very different from the ferromagnetic 
case.\cite{Loison.Simon.J1J2}
\item[2.] The behavior of $\chi_2^M$ as function of $U^M$ 
(Fig.~\ref{X2.1.KT}-\ref{X2.3.KT}) which give $\eta^M$.
\item[3.] The Universality of our results for $\eta^M$ for different models
(see table~\ref{table3}).
\item[4.] The dynamical properties shown in Fig.~\ref{fig.a3} are an another
indication that our choice for critical temperature is correct. 
Moreover the exponent
$\eta^M$ is in good agreement with those found in equilibrium properties.
\end{itemize}

Since the two calculations (in the HT and FSS regions) 
seem  correct we suggest an hypothesis similar to those
given for the Ising symmetry, i.e. different behavior in the 
FSS and HT regions.
We give hereafter a physical interpretation of this behavior.
At high temperature the radius of the vortex are very small and they are 
slightly connected together while the correlation length
for the Ising symmetry is bigger than the radius of the vortex. The 
situation is different in the FFS region. In this case the radius of the
vortex are large enough and the properties should depend on the domain walls
due to the Ising symmetry with the Ising correlation length $\xi_I$ 
smaller or of the
same size as the radius of the vortex. This argument is exactly the opposite
of those given to explain the difference of behavior for the Ising transition:
the "true" ferromagnetic Ising behavior appears only when $\xi_I$ is much
larger than a screening length $\lambda$, i.e. the $\xi_{KT}$. 
This difference of argumentation reflects the differences between the two 
symmetries: the Kosterlitz-Thouless transition is a transition 
driven by topological defects (vortex-antivortex), i.e. by "local" behavior, 
contrary to the Ising transition.  
We call our picture "quasi Universality class" in a similar way as
the Ising symmetry. The two interpretations diverge in the sense
that in the thermodynamic limit (infinite size) the new Universality 
class for the Ising symmetry could be not stable 
(however see Ref. \cite{Loison.XY.Potts})
while it is for the
Kosterlitz-Thouless transition. However the two behaviors are unstable 
in the high temperature region, i.e. sufficiently far away from 
the transitions.

\section{Conclusion}
This article is devoted to the study of the triangular 
frustrated $XY$ spin system in two dimensions
by extensive Monte Carlo simulations.

This system is characterized by two symmetries: a $U(1)$ symmetry due to 
the continuous nature of the $XY$ spins and an Ising symmetry due to the 
degeneracy of the ground state. Our study is restricted to the region 
where the system sizes that we simulate are smaller than the values of 
the correlation lengths of the $U(1)$ symmetry like those the 
Ising symmetry should have in an infinite system.

We have shown that the Kosterlitz-Thouless transition connected to the 
$U(1)$ symmetry appears at lower temperature than the Ising symmetry.
Our very large statistics (two order larger than previous
studies) allow us to take into account the corrections to the
scaling laws. With the knowledge of corrections we have shown 
that the Ising transition belongs to a new  Universality class, i.e.
the renormalization group flow tends toward a new "stable" fixed point. 
These corrections explain also the lack of Universality found by several 
authors. Knowing that the transition has a different behavior in the high
temperature phase we have introduced the idea of an "quasi fixed point"
or "quasi Universality class". 
In this interpretation  this new fixed point could be unstable in the 
thermodynamic limit but the true behavior cannot be reached in our limited 
accessible sizes in numerical simulations.
Accepting this interpretation, the situation is similar 
to those who appear in frustrated systems in three 
dimensions.\cite{LoisonSchotteHei}
We note that our interpretation goes further
than the interpretation of Olson\cite{Olson95} who predicts just an 
impossibility to obtain the true Ising behavior in the finite size scaling
region.

We have also studied the Kosterlitz-Thouless transition associated to 
the $U(1)$ symmetry. We have shown that this transition belongs to a new 
Universality class characterized by an universal exponent $\eta\sim0.36$
and a helicity jump smaller than those in the ferromagnetic case in 
disagreement with theoretical predictions.
We have introduced the concept of an "quasi Universality class"
for this transition to explain the 
different behaviors observed between the finite size scaling
and the high temperature regions. 
In this interpretation the new behavior is stable in the thermodynamic 
limit near the critical temperature but not when the temperature 
is much larger than the critical temperature.  
We have given a physical interpretation of this new concept.

To bring new informations and verify our predictions for these systems 
we have three possible ways. The first way is to improve the theoretical
approaches of these systems. In particular it should be extremely
interesting to have a theoretical result for exponents of the new
fixed point for the Ising symmetry. However since it is  difficult to 
describe the coupling between the two symmetries, the precise 
calculation of exponents seems, for the moment, an impossible task. 
Another possibility is the experimental approach in the
Josephson-junction array\cite{Lobb88,Ling96} or in films of 
$^3$He.\cite{Halsey85,Korshunov86,Kotsubo87,Xu90}  
However one of 
the problems of experiments is to have access to the interesting quantities. For
example it is usually difficult to measure the chirality in these systems
because it is not related directly to measurable quantities. 
The easiest way to obtain reliable information seem to be numerical
simulations. 
Indeed the use of powerful computers and fast 
algorithms allows to obtain precise results out of reach some years ago.     
We have to study different models to observe the Universality class
in the finite size scaling region but also in the high temperature
region to verify the picture predicted in this work.

\section {Acknowledgments}
This work was supported by the Alexander von Humboldt Foundation. 
We are grateful to Professors K. D. Schotte and I. Peschel for 
discussions and to L. Beierlein for critically reading the manuscript.

\newpage

\begin{table}[t]
\begin{center}
\begin{tabular}{c|c|c|c|c|c}
\hspace{-10pt}
\begin{tabular}{c}$L$ \end{tabular}
\hspace{-10pt}
&
\begin{tabular}{c}$t_{th}$\end{tabular}
\hspace{-10pt}
&
\begin{tabular}{c}$t_{av}$\end{tabular}
\hspace{-10pt}
&
\begin{tabular}{c}$\tau_\kappa$\end{tabular}
\hspace{-10pt}
&
\begin{tabular}{c}$\tau_s$\end{tabular}
\hspace{-10pt}
&
\begin{tabular}{c}${t_{av} \over 2\tau_\kappa+\tau_s}$\end{tabular}
\hspace{-10pt}
\\
\hline
12& $3. 10^5$ & $30. 10^6$&11.9(1)&100&$2.4\,10^5$
\hspace{-10pt}
\\
\hline
18& $3. 10^5$ & $30. 10^6$&13.0(6)&100&$2.4\,10^5$
\hspace{-10pt}
\\
\hline
24& $5. 10^5$ & $30. 10^6$&26(1)&100&$2.0\,10^5$
\hspace{-10pt}
\\
\hline
36& $5. 10^5$ & $20. 10^6$&50(2)&100&$1.0\,10^5$
\hspace{-10pt}
\\
\hline
48& $5. 10^5$ & $18. 10^6$&73.(3)&100&$7.3\,10^4$
\hspace{-10pt}
\\
\hline
60& $7. 10^5$ & $30. 10^6$&98(5)&100&$1.0\,10^5$
\hspace{-10pt}
\\
\hline
81& $7. 10^5$ & $30. 10^6$&163(13)&163&$6.1\,10^4$
\hspace{-10pt}
\\
\hline
105& $1. 10^6$ & $27. 10^6$&236(18)&236&$3.8\,10^4$
\hspace{-10pt}
\\
\hline
123& $1. 10^6$ & $21. 10^6$&300(28)&300&$2.3\,10^4$
\hspace{-10pt}
\\
\hline
150& $1. 10^6$ & $17. 10^6$&404(66)&404&$1.4\,10^4$
\hspace{-10pt}
\end{tabular}
\end{center}
\caption{\protect
\label{table.tau}
}
\end{table}

\begin{table}[t]
\addcontentsline{toc}{part}{Table}
\begin{center}
\begin{tabular}{c|c|c|c|c|c|c|c|c}
\hspace{-10pt}
\begin{tabular}{c}system \ \ \end{tabular}
\hspace{-10pt}
&
\begin{tabular}{c}ref.\end{tabular}
\hspace{-10pt}
&
\begin{tabular}{c}$method$\end{tabular}
\hspace{-10pt}
&
\begin{tabular}{c}$L_{max}$\end{tabular}
\hspace{-10pt}
&
\begin{tabular}{c}$N_{MC}$\end{tabular}
\hspace{-10pt}
&
\begin{tabular}{c}$\tau_{max}$\end{tabular}
\hspace{-10pt}
&
\begin{tabular}{c}$t_{ind}$\end{tabular}
\hspace{-10pt}
&
\begin{tabular}{c}$\nu_I$\end{tabular}
\hspace{-10pt}
&
\begin{tabular}{c}$\eta_I$\end{tabular}
\hspace{-10pt}
\\
\hline
FF$XY$&\cite{Nicolaides91}&MC HT&128&&&&1.009(26)&0.274(20)
\hspace{-10pt}
\\
\hline
FF$XY$&\cite{Jose96}&MC HT&128&&&&0.898(3)&
\hspace{-10pt}
\\
\hline
FF$XY$&\cite{Thijssen90}&MC TM&12&&&&$\sim$1&0.400(20)
\hspace{-10pt}
\\
\hline
FF$XY$&\cite{Granato93}&MC TM&12&&&&0.800(50)&0.380(20)
\hspace{-10pt}
\\
\hline
FF$XY$&\cite{LeeLee94}&MC Micro FSS&64&&&&0.813(5)&0.219(18)
\hspace{-10pt}
\\
\hline
FF$XY$&\cite{Luo98}&MC dyn.&256&&&&0.810(20)&0.262(6)
\hspace{-10pt}
\\
\hline
FF$XY$&\cite{JLee91}&MC FSS&40&5.10$^6$&386$^{(a)}$&6500&0.850(30)&0.310(30)
\hspace{-10pt}
\\
\hline
Zig-Zag$_1$&\cite{Benakly97}&MC FSS&36&12.10$^6$&310$^{(a)}$&20000&0.800(10)&0.290(20)
\hspace{-10pt}
\\
\hline
Zig-Zag$_2$&\cite{Benakly97}&MC FSS&36&12.10$^6$&310$^{(a)}$&20000&0.780(20)&0.320(40)
\hspace{-10pt}
\\
\hline
Zig-Zag$_3$&\cite{Boubcheur98}&MC FSS&140&3.10$^6$&7515$^{(a)}$&200&0.852(2)&0.203(1)
\hspace{-10pt}
\\
\hline
TA&\cite{LeeLee98}&MC Micro FSS&60&&&&0.830(10)&0.250(20)
\hspace{-10pt}
\\
\hline
TA&\cite{JLee91}&MC FSS&30&5.10$^6$&494$^{(b)}$&5000&0.830(40)&0.280(40)
\hspace{-10pt}
\\
\hline
TA &\cite{Caprioti97}&MC FSS&120&6.10$^4$&2000$^{(b)}$&15&$\sim$1&$\sim$0.25
\hspace{-10pt}
\\
\hline
TA &this work&MC FSS&150&17.10$^6$&404&14000&0.815(20)&0.227(4)
\hspace{-10pt}
\\
\hline
TA &this work&MC dyn&300&&&&0.818(9)&0.235(5)
\hspace{-10pt}
\\
\hline
$J_1-J_2$&\cite{Fernandez91}&MC FSS &150&5.10$^5$&10000$^{(c)}$&25&0.900(200)&0.200(500)
\hspace{-10pt}
\\
\hline
$J_1-J_2$&\cite{Loison.Simon.J1J2}&MC FSS &150&32.10$^6$&50&32000&0.795(20)&0.250(5)
\hspace{-10pt}
\\
\hline
$XY$-Ising$_1$&\cite{Nightingale95}&MC TM&30&&&&0.79&0.40
\hspace{-10pt}
\\
\hline
$XY$-Ising$_2$&\cite{Nightingale95}&MC TM&30&&&&0.66&0.40
\hspace{-10pt}
\\
\hline
$XY$-Ising$_3$&\cite{Granato91}&MC FSS&40&5.10$^6$&?&?&0.76-0.86&0.24-0.42
\hspace{-10pt}
\\
\hline
Villain&\cite{Olson95}&MC HT &256&&&&1.02(5)&
\hspace{-10pt}
\\
\hline
19-vertex&\cite{Knops94}&MC TM&15&&&&0.770(30)&0.280(20)
\hspace{-10pt}
\\
\hline
1D-quantum&\cite{Granato92}&MC TM&14&&&&0.810(40)&0.470(40)
\hspace{-10pt}
\\
\hline
CG&\cite{JRLee94}&MC FSS&30&10$^6$&&&0.840(50)&0.260(40)
\hspace{-10pt}
\\
\hline
RSOS&\cite{LeeLeeRSOS}&MC Micro&22&&&&$\sim1$&0.25
\hspace{-10pt}
\end{tabular}
\end{center}
\caption{\label{table.revue.I}
$^{(a)}$estimated  using Ref. \protect\cite{Pawig98};
$^{(b)}$estimated  using this work;
$^{(c)}$estimated  using Ref. \protect\cite{Loison.Simon.J1J2}.
}
\end{table}

\begin{table}[t]
\begin{center}
\begin{tabular}{c|c|c|c|c|c|c|c}
\hspace{-10pt}
\begin{tabular}{c}symmetry \end{tabular}
\hspace{-10pt}
&
\begin{tabular}{c}$T_c$ \end{tabular}
\hspace{-10pt}
&
\begin{tabular}{c}$U_*$ \end{tabular}
\hspace{-10pt}
&
\begin{tabular}{c}$\nu$ \end{tabular}
\hspace{-10pt}
&
\begin{tabular}{c}$\gamma$ \end{tabular}
\hspace{-10pt}
&
\begin{tabular}{c}$\eta$ \end{tabular}
\hspace{-10pt}
&
\begin{tabular}{c}$\beta$\end{tabular}
\hspace{-10pt}
&
\begin{tabular}{c}z\end{tabular}
\hspace{-10pt}
\\
\hline
$Z_2$ (eq)&0.5122(1)&0.6320(20)&0.815(20)&1.450(42)&0.227(9)$^a$&0.0896(71)&2.30(4)
\hspace{-10pt}
\\
\hline
$Z_2$ (dyn)&&&0.818(9)&1.445(20)&0.235(7)$^a$&0.0967(13)&2.39(5)
\hspace{-10pt}
\\
\hline
$Z_2^{J_1-J_2}$ (eq) &0.56465(8)&0.6269(7)&0.795(20)&1.391(43)&0.250(10)$^a$&0.101(10)&2.29(4)
\hspace{-10pt}
\\
\hline
$U(1)$ (eq)&0.5102(1)&0.6497(12)&& &0.365(10)&&
\hspace{-10pt}
\\
\hline
$U(1)$ (dyn)&&&& &0.36&&2.10
\hspace{-10pt}
\\
\hline
$U(1)^{J_1-J_2}$ (eq)&0.56271(5)&0.638(5)&& &0.345(5)&&
\hspace{-10pt}
\end{tabular}
\end{center}
\caption{\protect\label{table3}
$^a$Calculated using $2-\eta=\gamma/\nu$.
}
\end{table}

\newpage
\begin{center}
TABLE CAPTIONS
\end{center}

Table~\ref{table.tau}:
Number of Monte Carlo steps to thermalize $t_{th}$ and to average
$t_{av}$ as function of the lattice size.
$\tau_\kappa$ are calculated with shorter MC runs.
We collect the data every $\tau_s$.
The last column gives the number of ''independent'' measurements.

Table~\ref{table.revue.I}:
Results for the two dimensional frustrated $XY$ systems with a breakdown
of symmetry $U(1) \otimes Z_2$.
Critical exponents are associated to the Ising symmetry.
I= Ising;
FF$XY$= Fully Frustrated $XY$;
CG= Coulomb gas;
TA= Triangular Antiferromagnetic;
MC= Monte Carlo; HT= High Temperatures; FSS= Finite Size Scaling;
Micro=Microcanonic TM= Transfer Matrix; dyn= dynamic;
RSRG= Real Space Renormalization Group; RSOS= Restricted Solid on Solid.
The indices $_1,_2,_3$ for the Zig-Zag or the $XY$-Ising model refer
to different choice of internal free parameter.
$t_{ind}=N_{MC}/2\tau_{max}$;

Table~\ref{table3}:
Summary of our results for the Ising symmetry ($Z_2$) and the
$XY$ symmetry $U(1)$. The results for the $J_1-J_2$ model is from
Ref. \protect\cite{Loison.Simon.J1J2}.
(eq)=equilibrium properties; (dyn)=short time dynamical properties.

\newpage
\begin{center}
FIGURE CAPTIONS
\end{center}

\begin{figure} 
\begin{center}

\psfig{figure=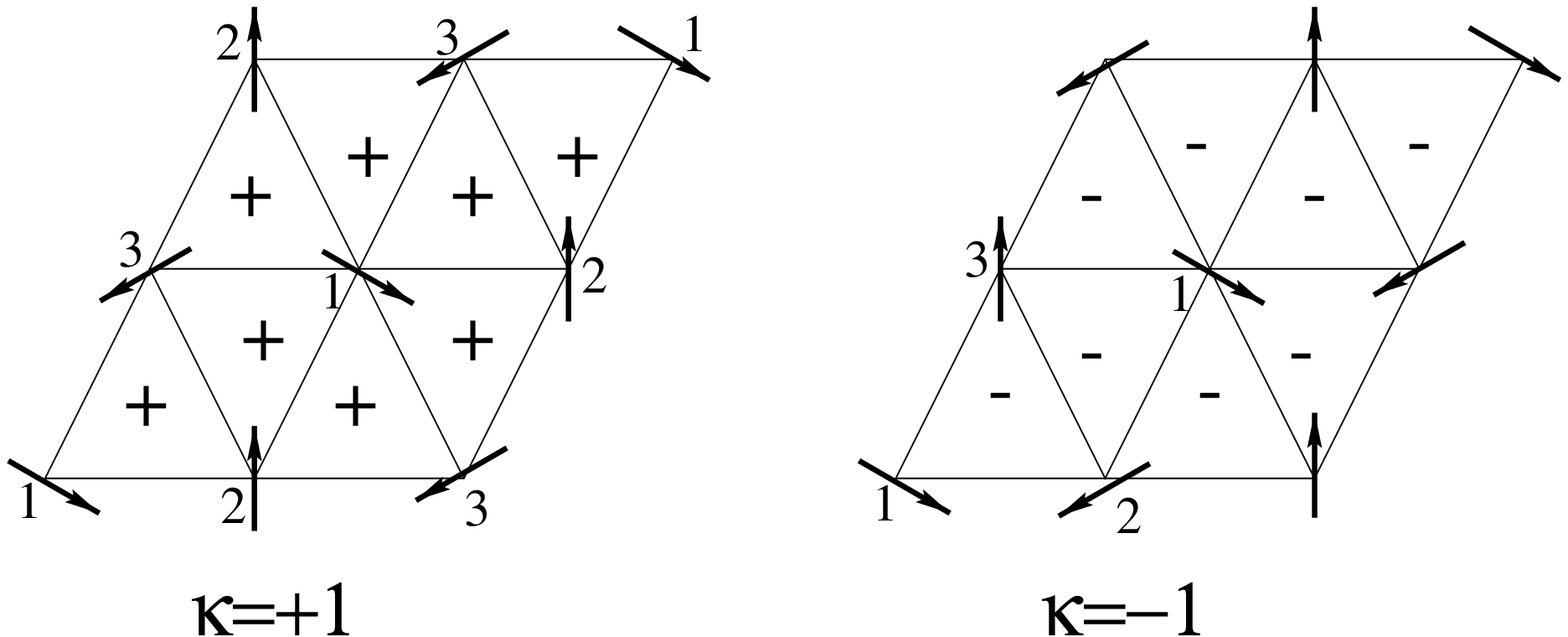,width=10.5cm}
\caption{
\label{fig.GS}
}
\vspace{1cm}

\psfig{figure=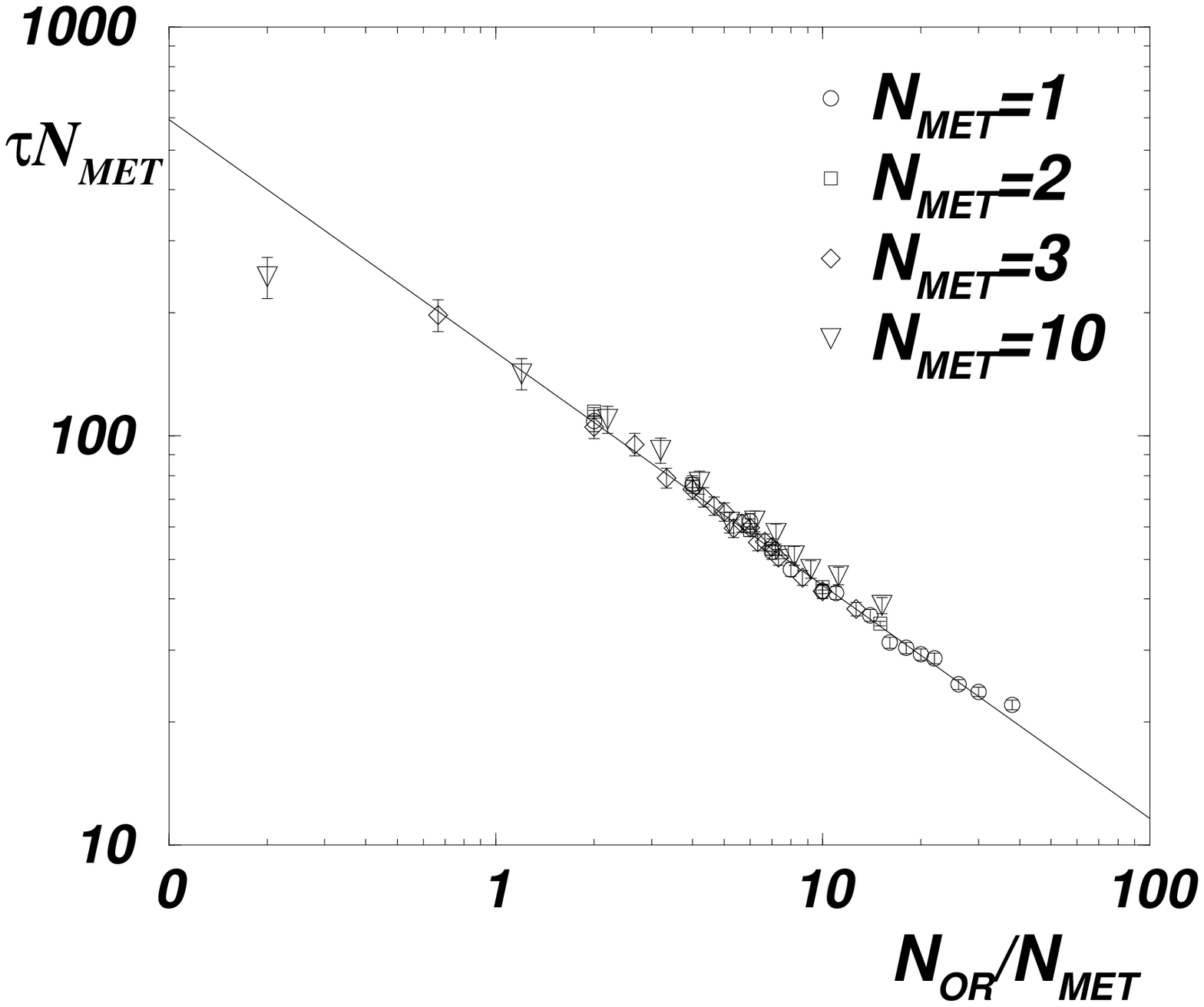,width=6.5cm}
\caption{
\label{fig.NMET.NOR}
}
\vspace{1cm}

\psfig{figure=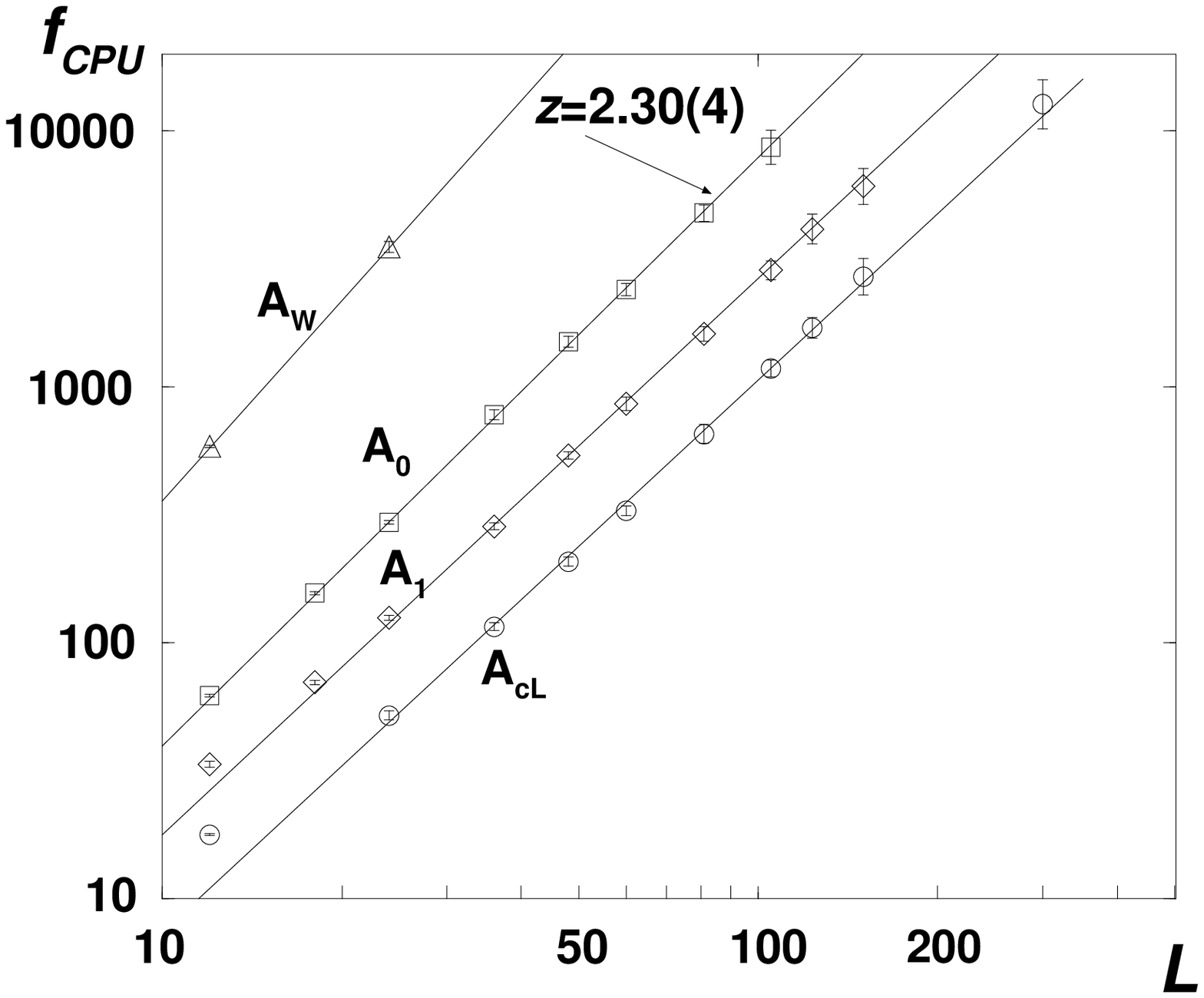,width=6.5cm}
\caption{
\label{fig.tau.L}
}
\vspace{1cm}

\psfig{figure=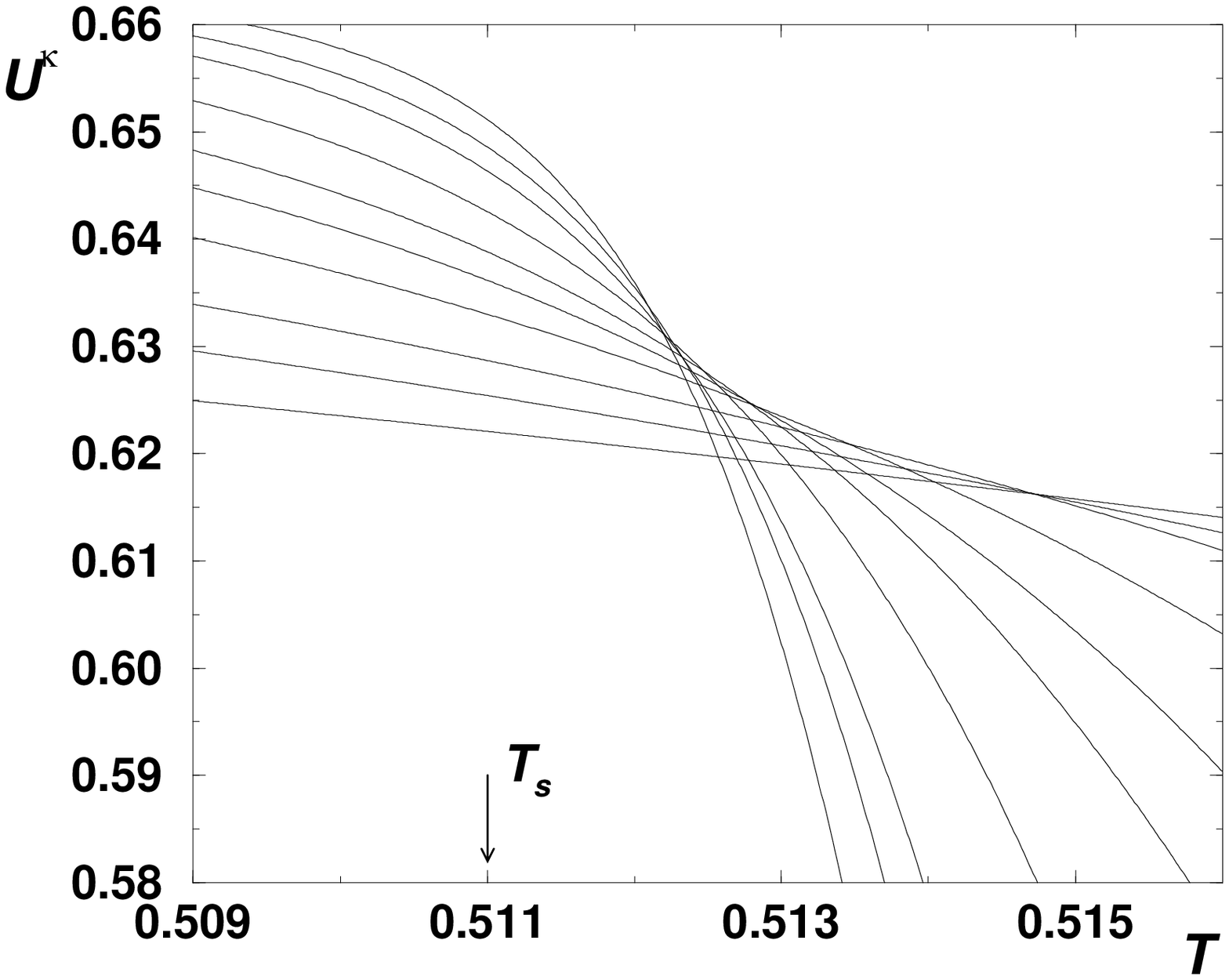,width=6.5cm}
\caption{
\label{figCM}
}
\vspace{1cm}

\psfig{figure=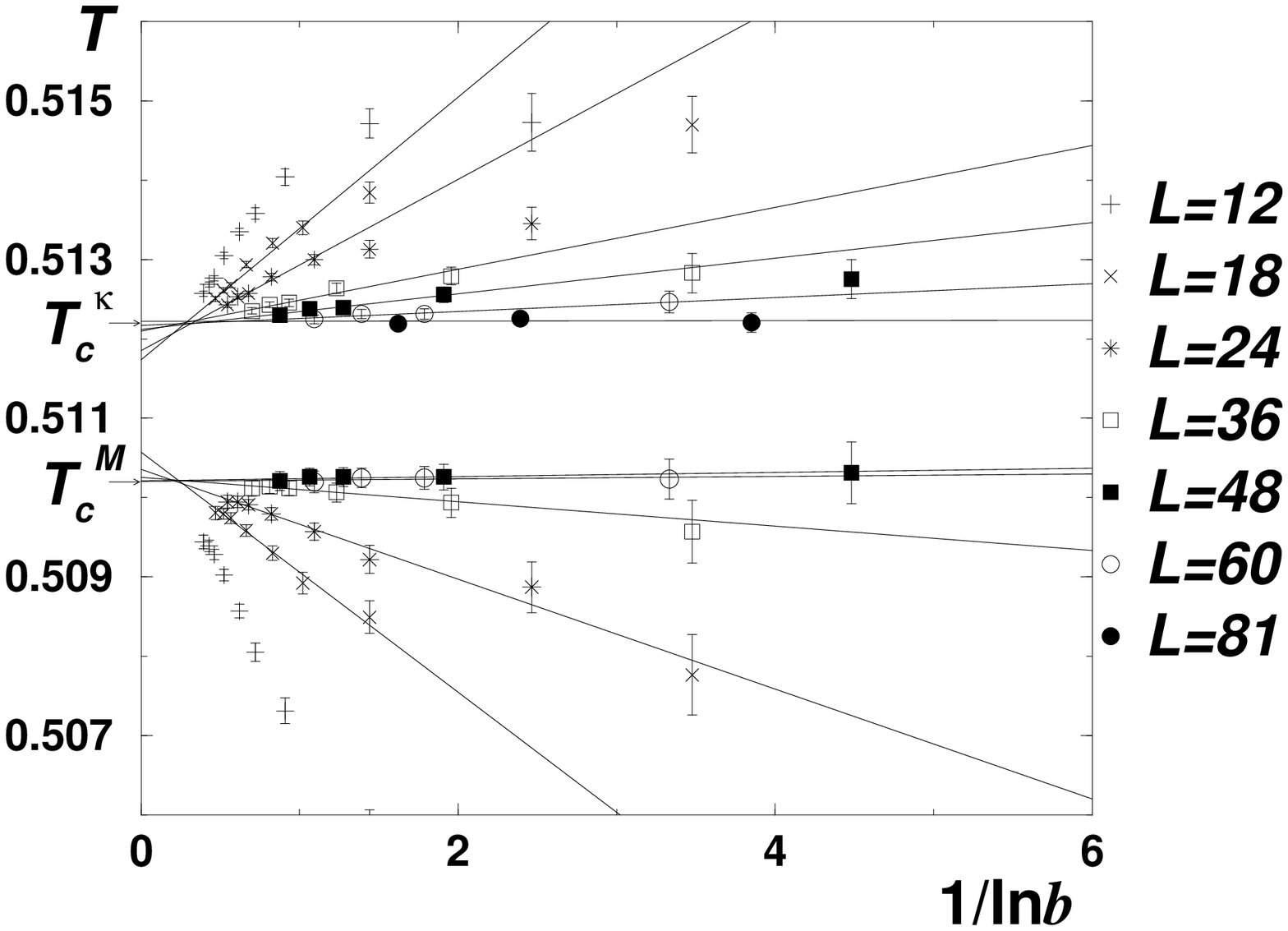,width=6.5cm}
\caption{
\label{figTc2.all}
}
\vspace{1cm}

\psfig{figure=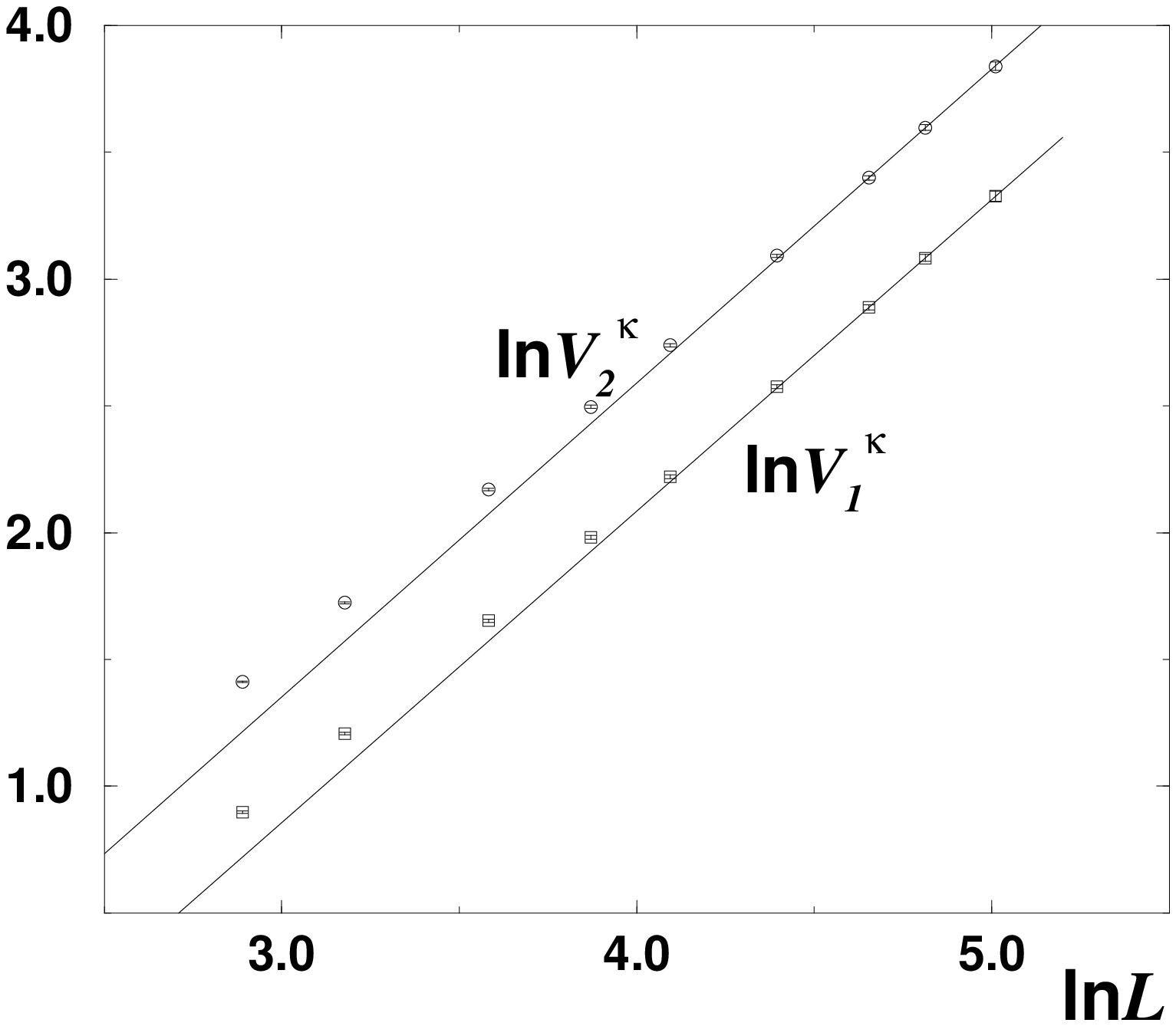,width=6.5cm}
\caption{
\label{figV}
}
\vspace{1cm}

\psfig{figure=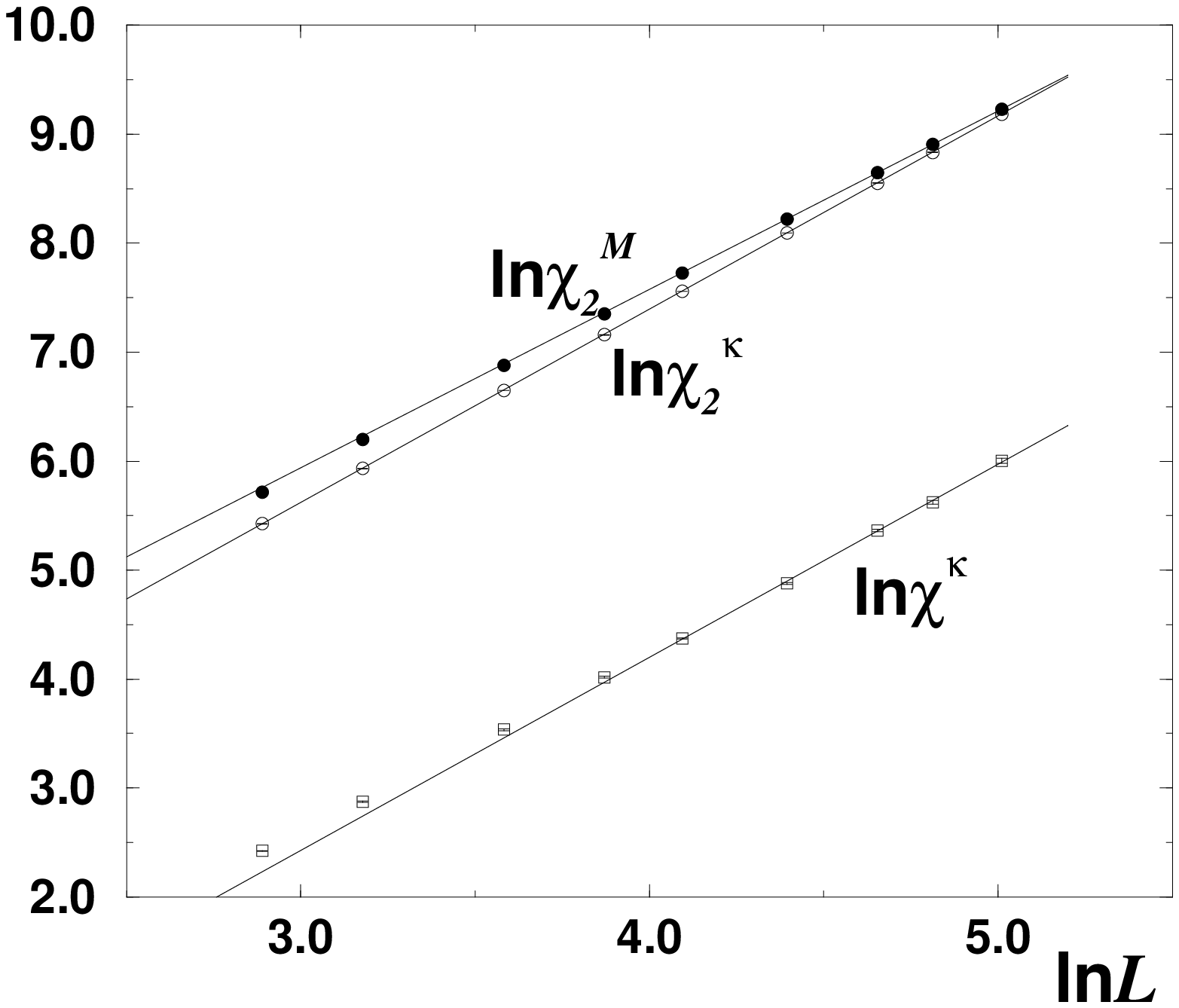,width=6.5cm}
\caption{
\label{figX.all}
}
\vspace{1cm}

\psfig{figure=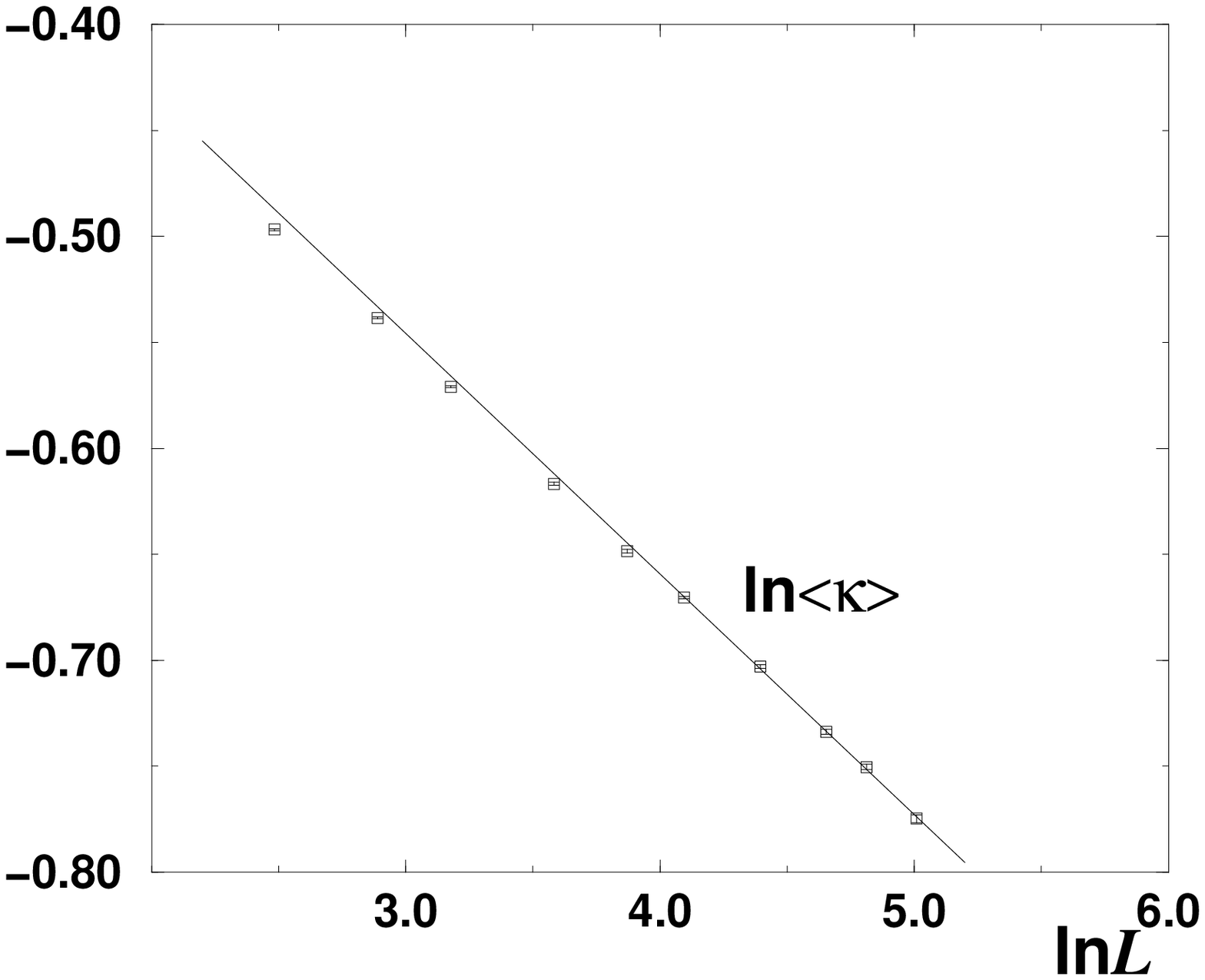,width=6.5cm}
\caption{
\label{figM}
}
\vspace{1cm}

\psfig{figure=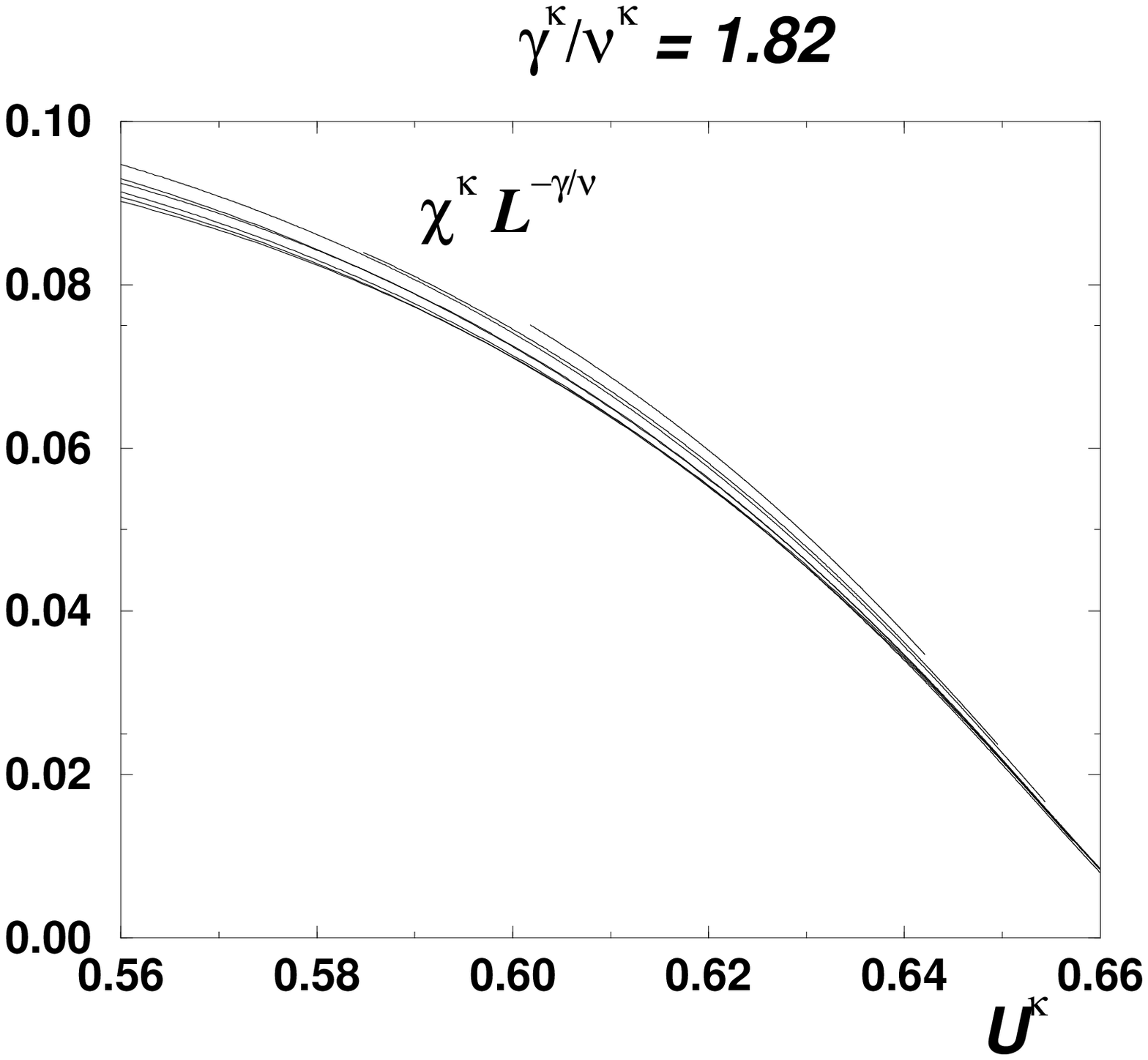,width=6.5cm}
\caption{
\label{X2.1}
}
\vspace{1cm}

\psfig{figure=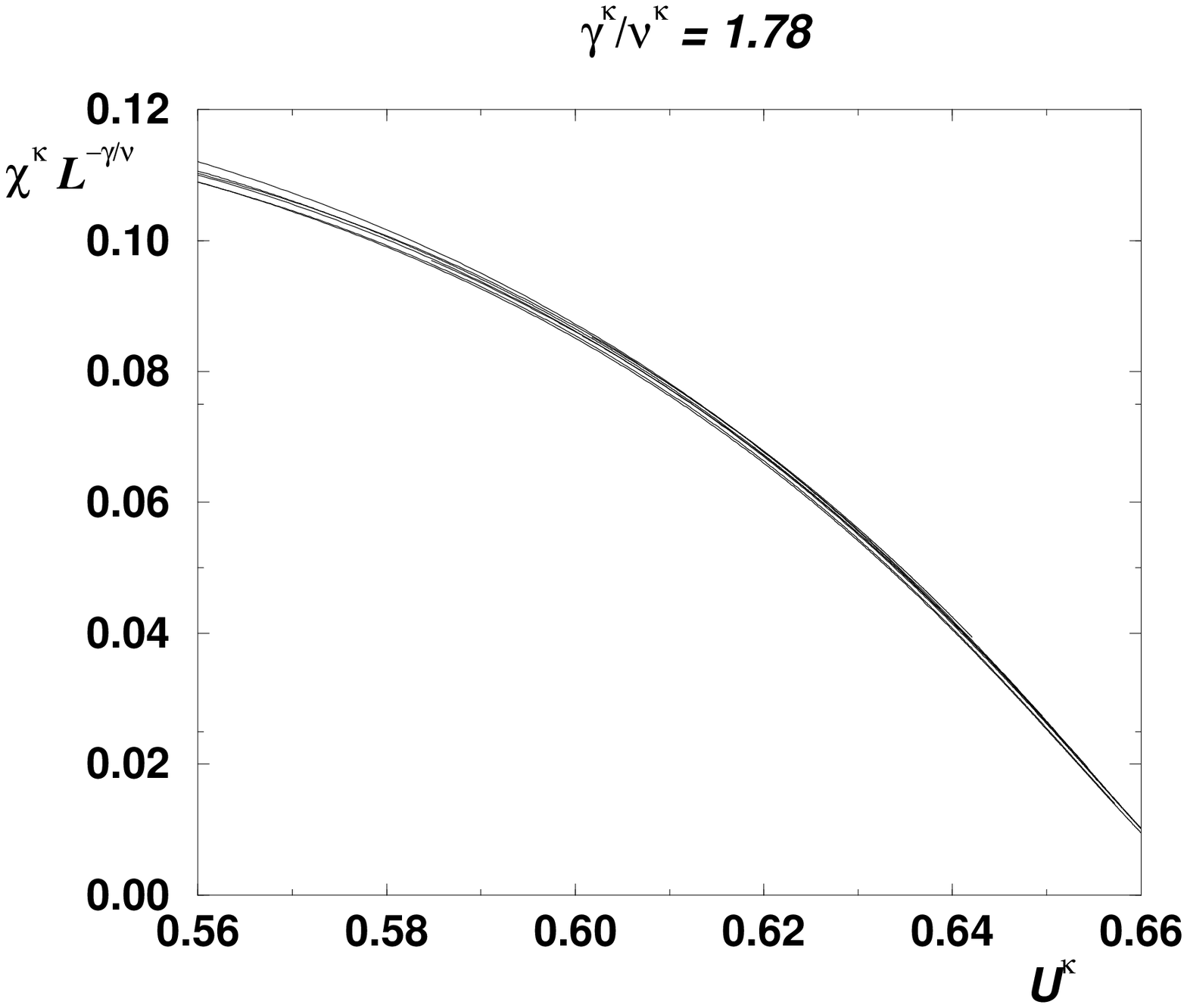,width=6.5cm}
\caption{
\label{X2.2}
}
\vspace{1cm}

\psfig{figure=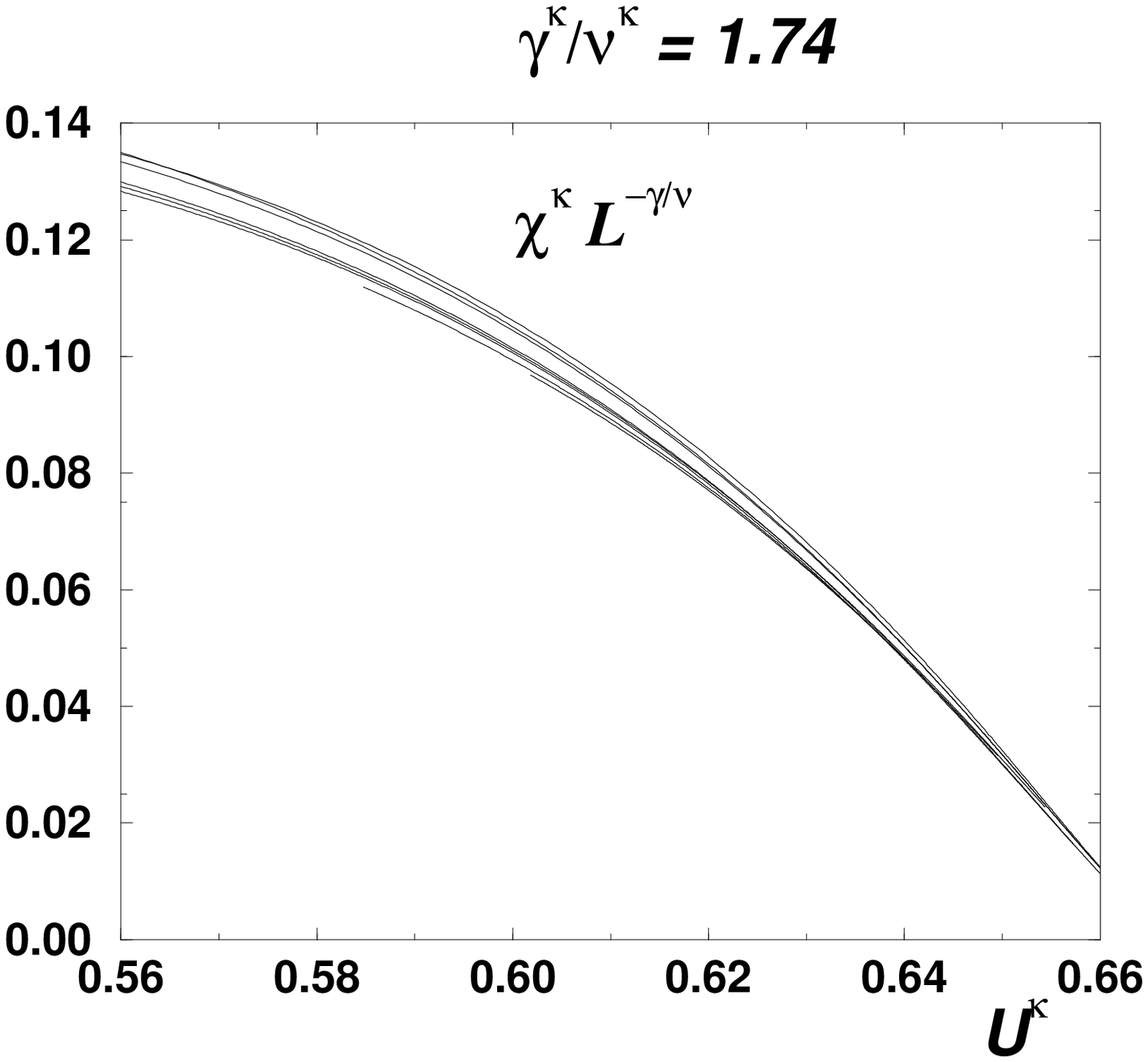,width=6.5cm}
\caption{
\label{X2.3}
}
\vspace{1cm}

\psfig{figure=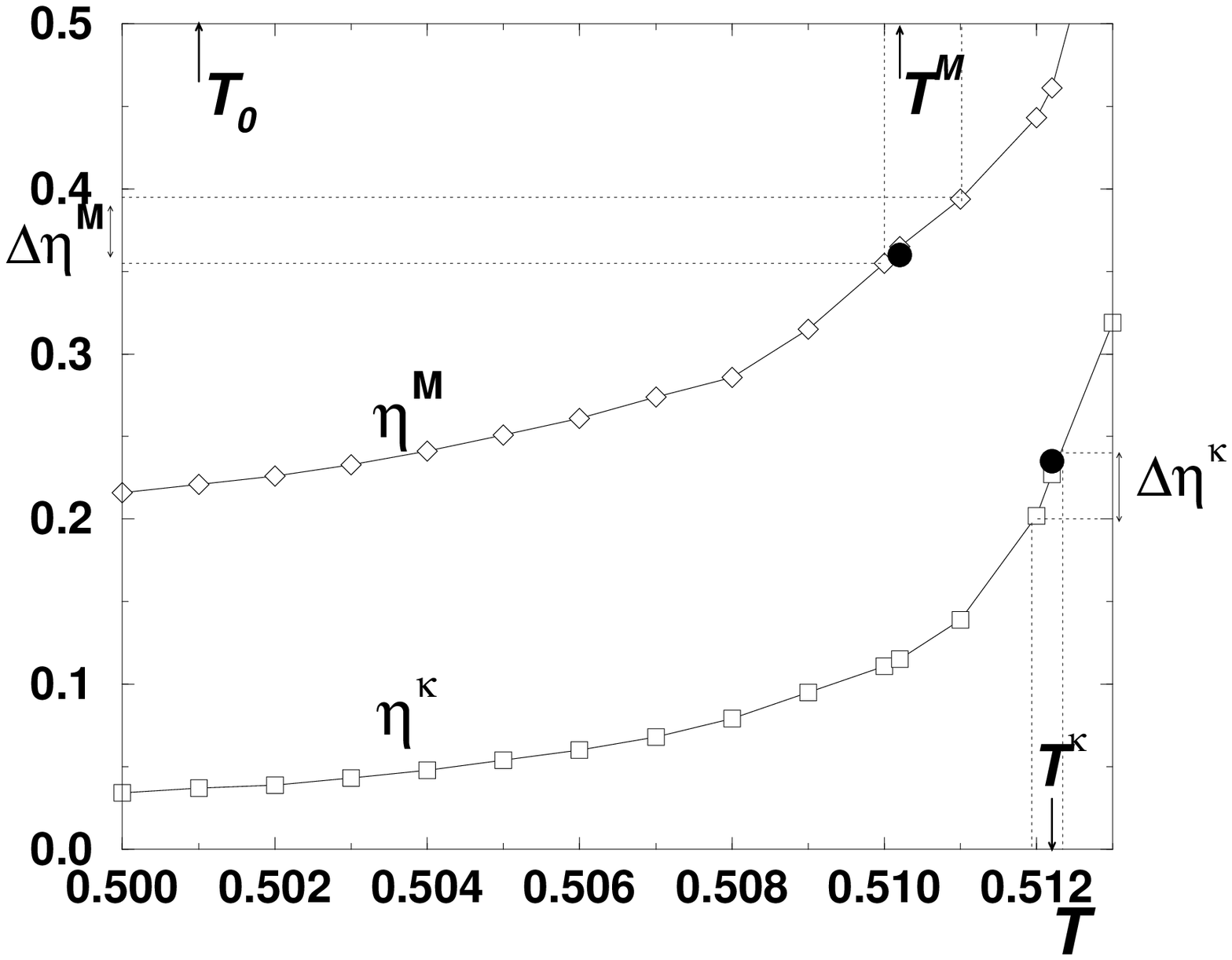,width=6.5cm}
\caption{
\label{fig.eta}
}
\vspace{1cm}

\psfig{figure=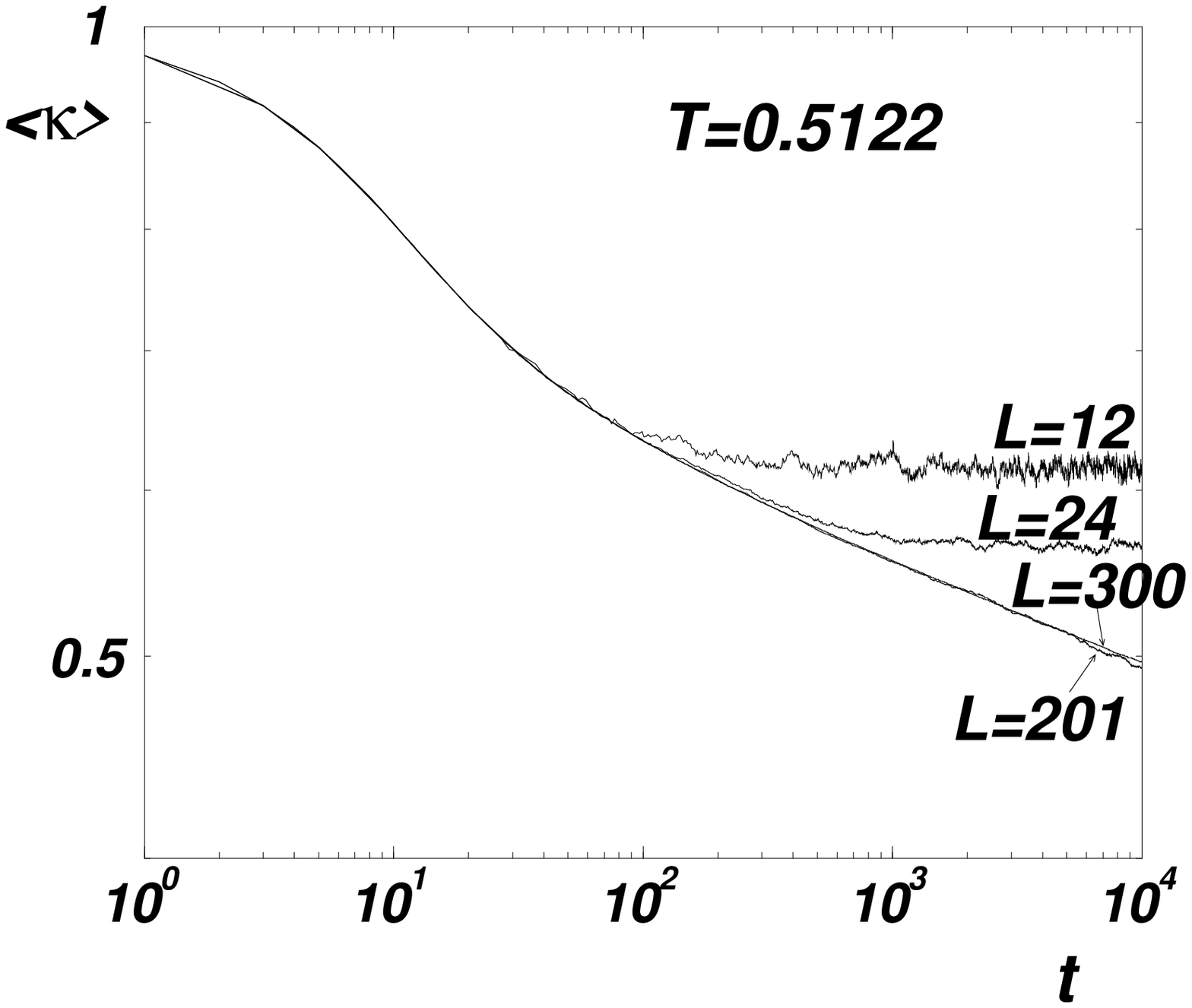,width=6.5cm}
\caption{
\label{fig.a4}
}
\vspace{1cm}

\psfig{figure=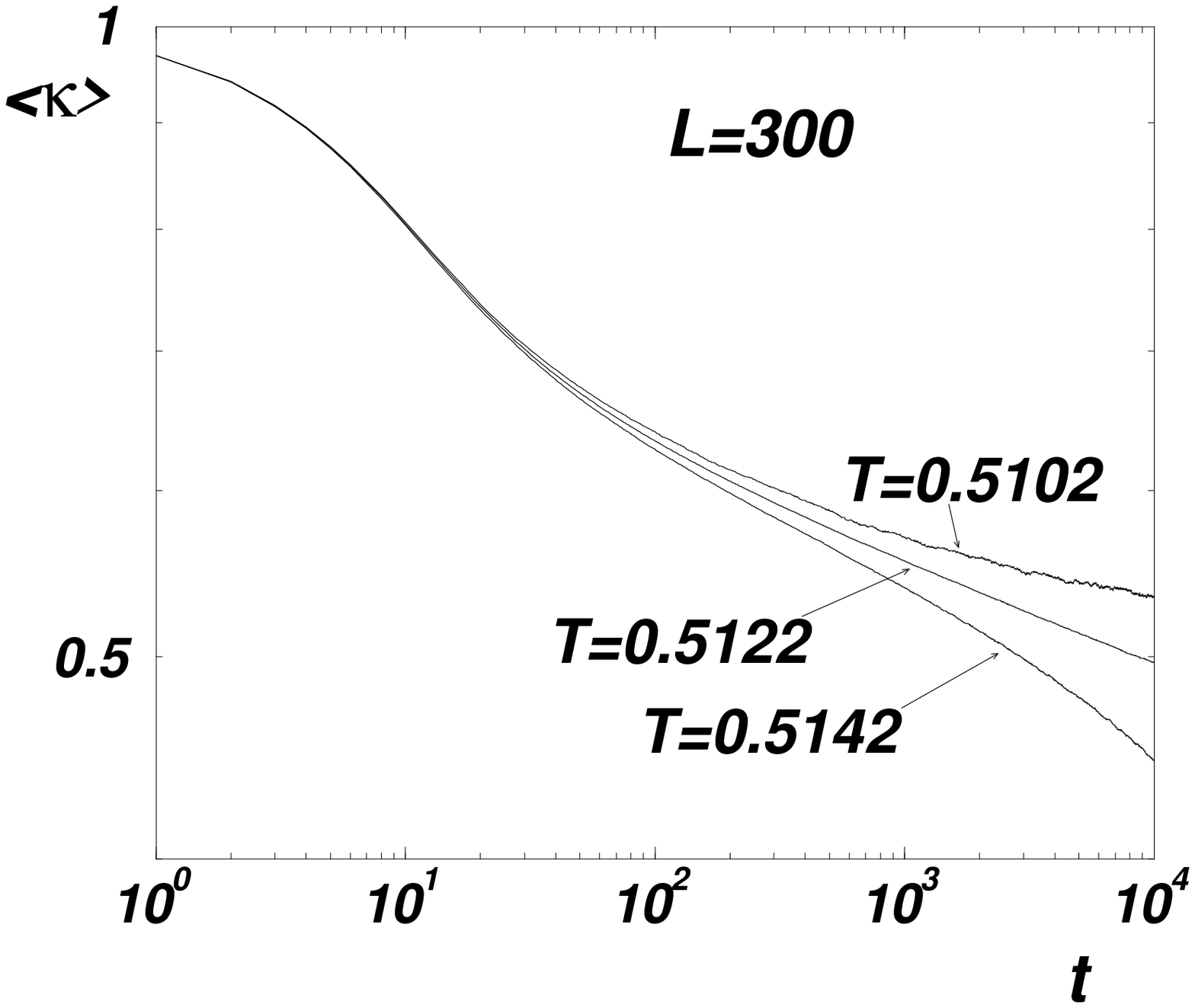,width=6.5cm}
\caption{
\label{fig.a2}
}
\vspace{1cm}

\psfig{figure=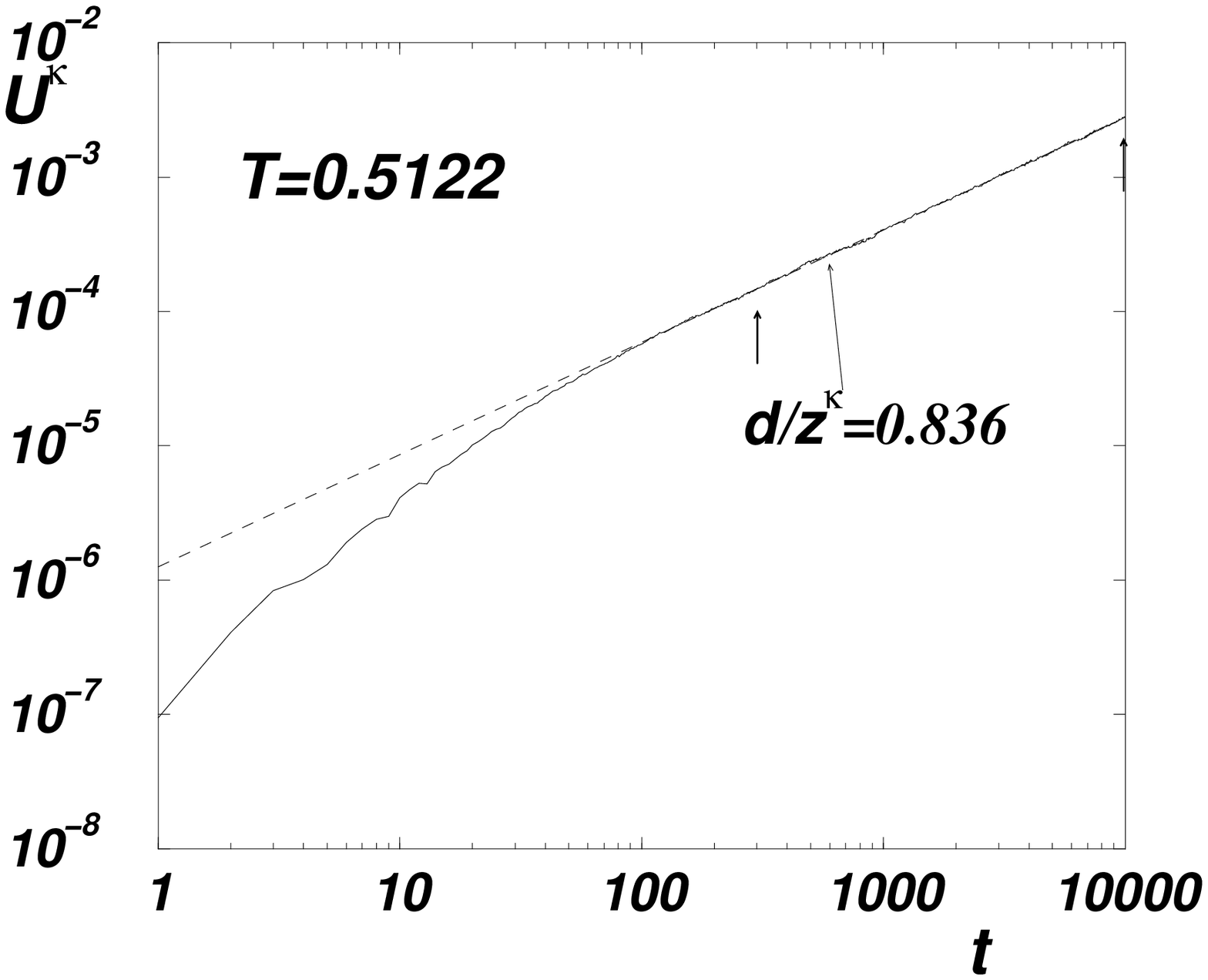,width=6.5cm}
\caption{
\label{fig.U.t}
}
\vspace{1cm}

\psfig{figure=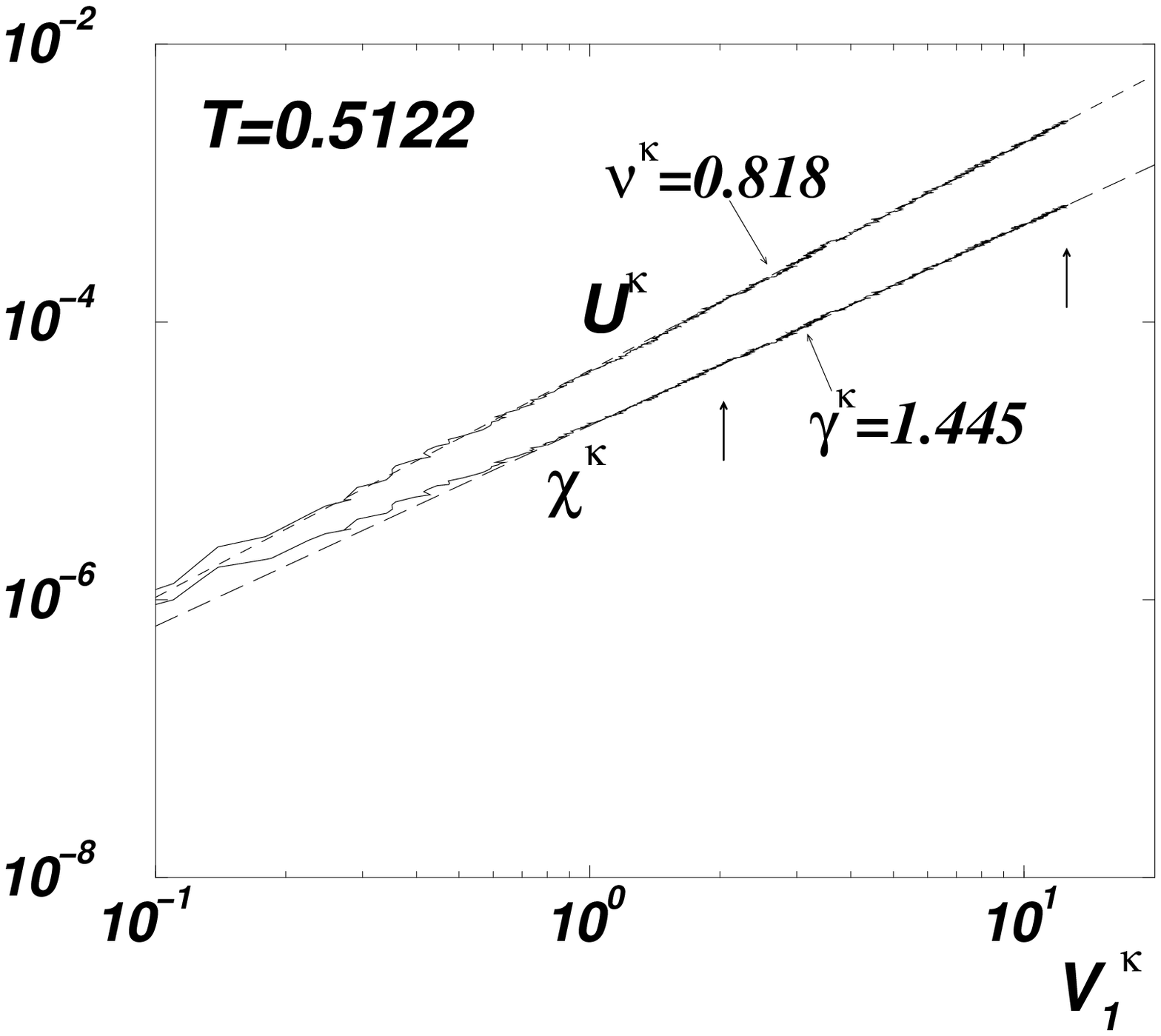,width=6.5cm}
\caption{
\label{fig.U.X.V1}
}
\vspace{1cm}

\psfig{figure=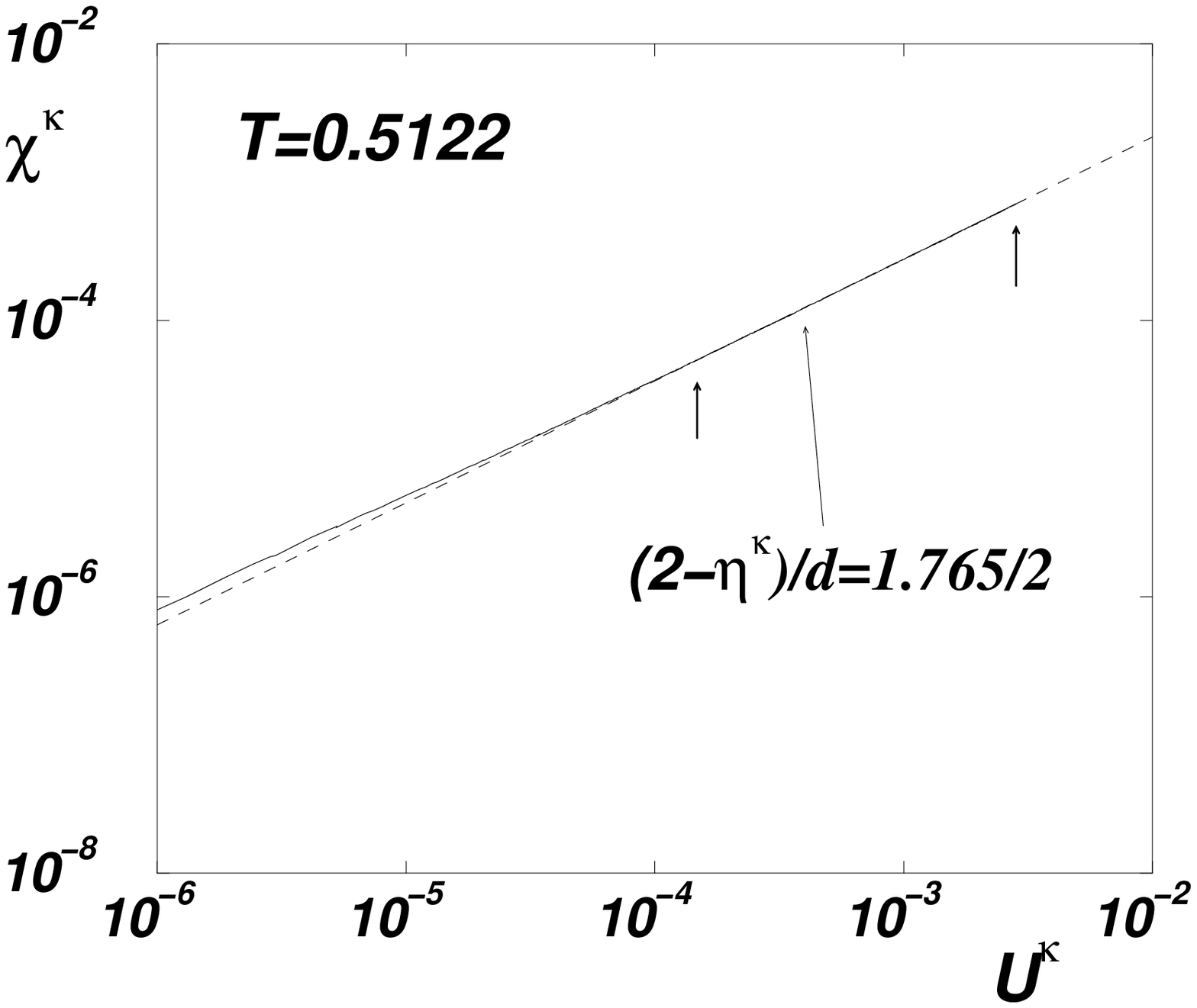,width=6.5cm}
\caption{
\label{fig.X.U}
}
\vspace{1cm}

\psfig{figure=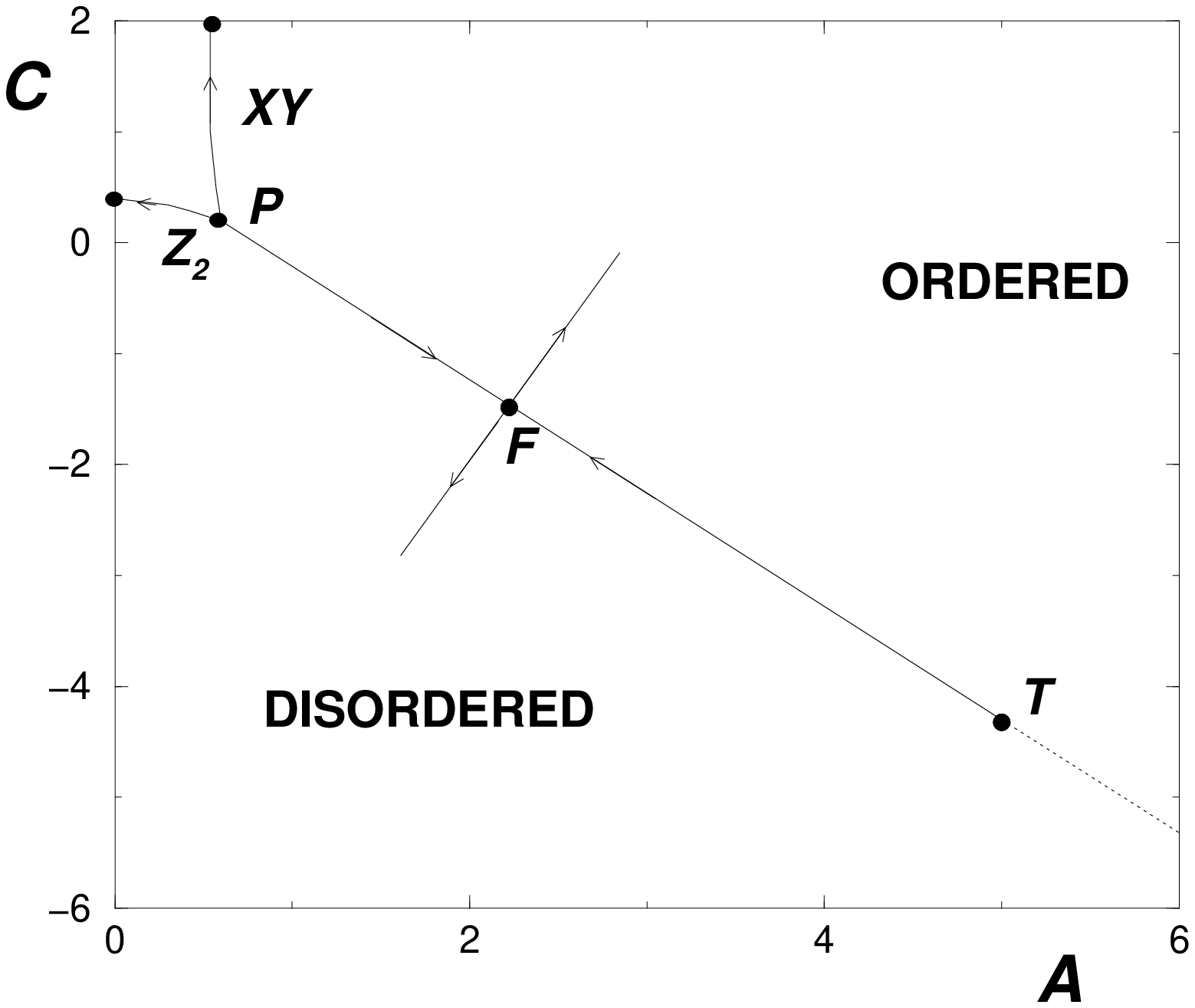,width=6.5cm}
\caption{
\label{fig.I-XY}
}
\vspace{1cm}

\psfig{figure=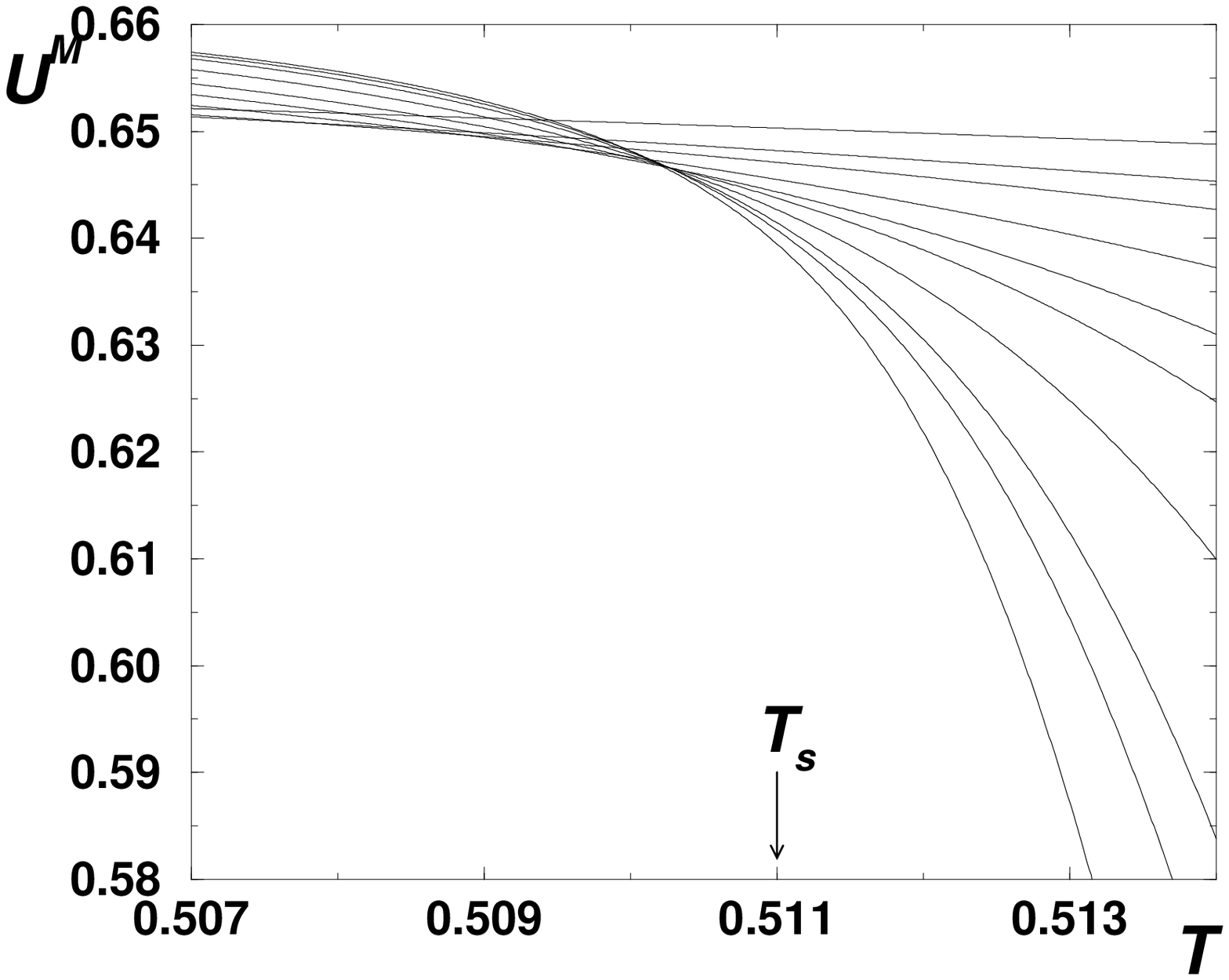,width=6.5cm}
\caption{
\label{figCM.KT}
}
\vspace{1cm}

\psfig{figure=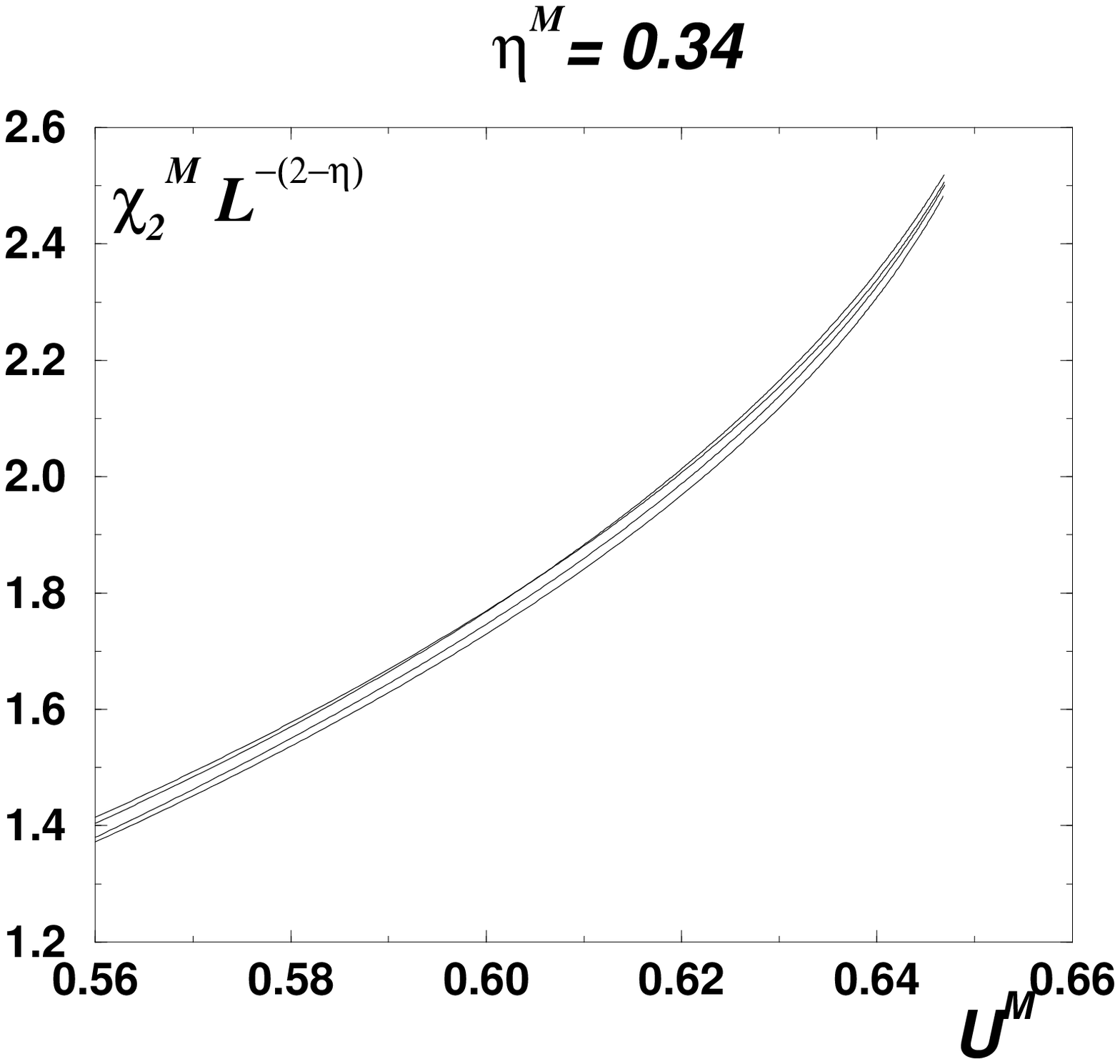,width=6.5cm}
\caption{
\label{X2.1.KT}
}
\vspace{1cm}

\psfig{figure=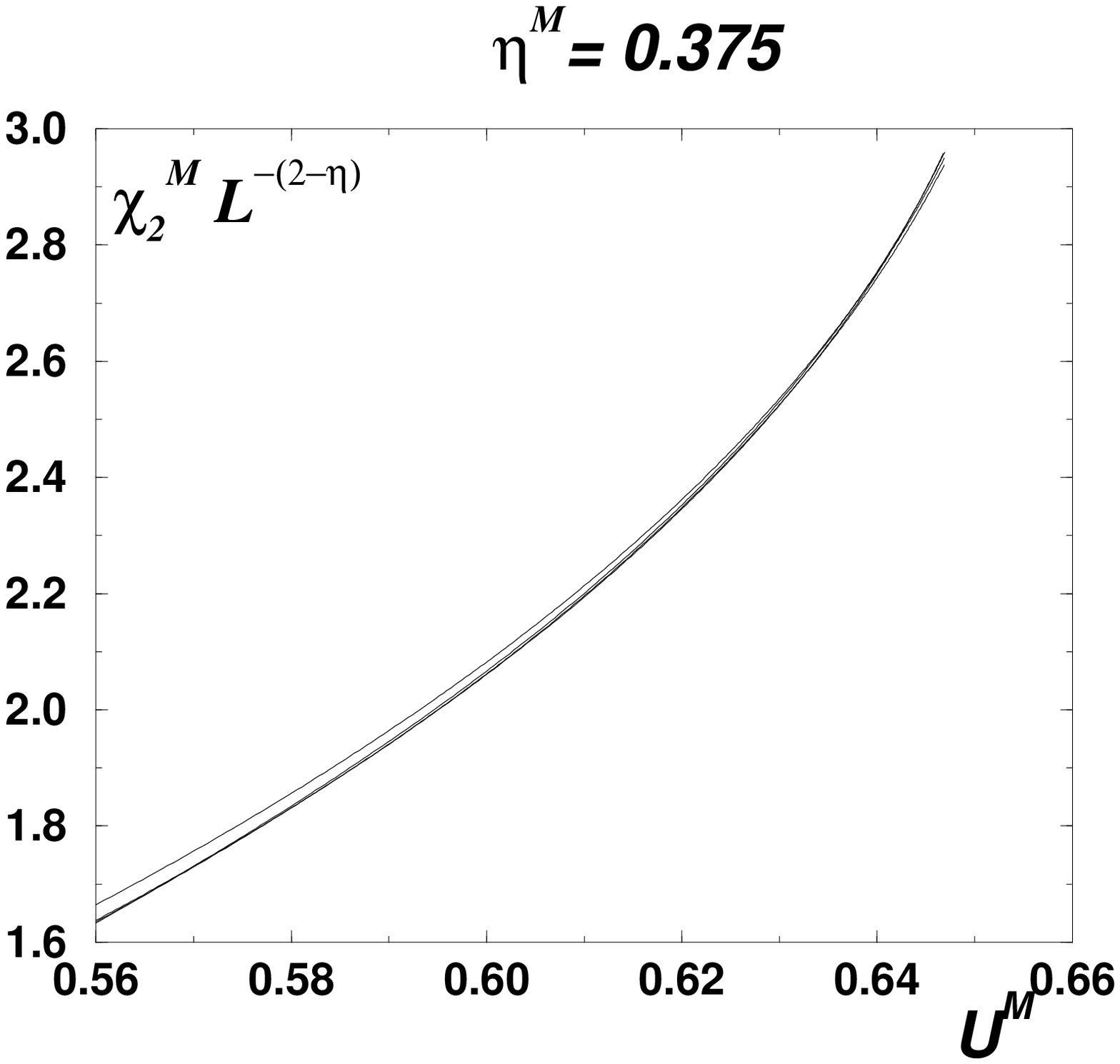,width=6.5cm}
\caption{
\label{X2.2.KT}
}
\vspace{1cm}

\psfig{figure=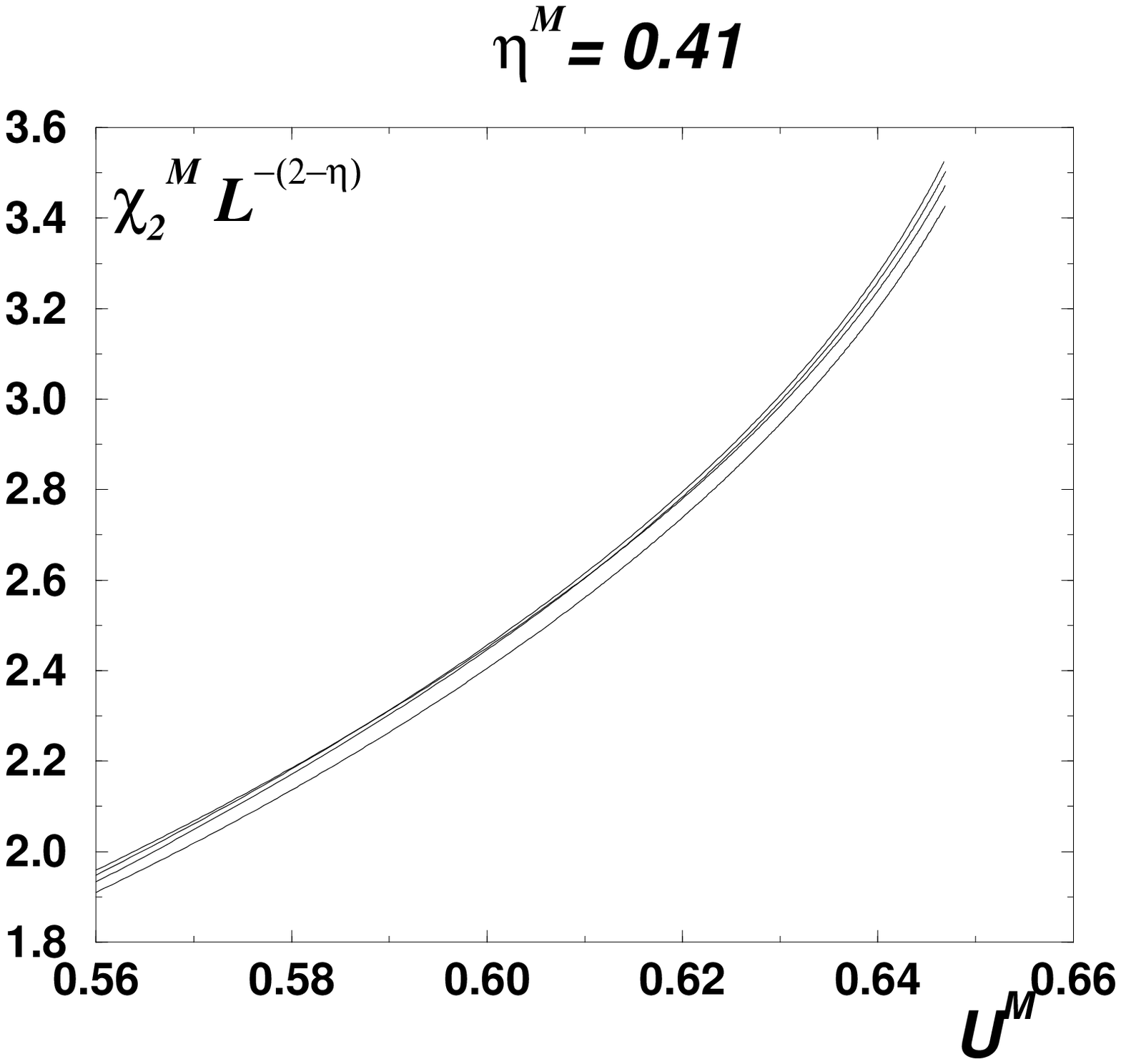,width=6.5cm}
\caption{
\label{X2.3.KT}
}
\vspace{1cm}

\psfig{figure=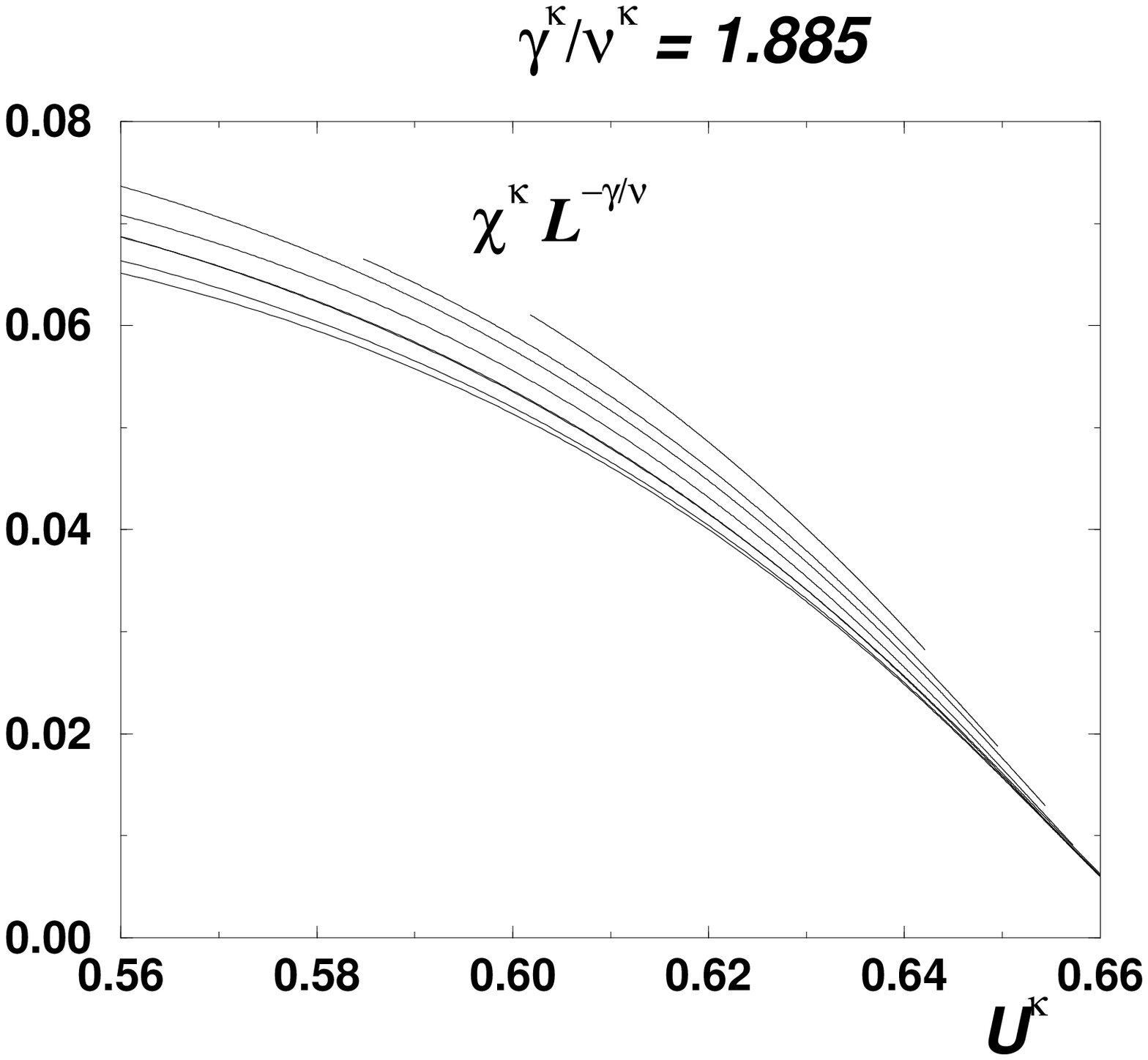,width=6.5cm}
\caption{
\label{X2.4}
}
\vspace{1cm}

\psfig{figure=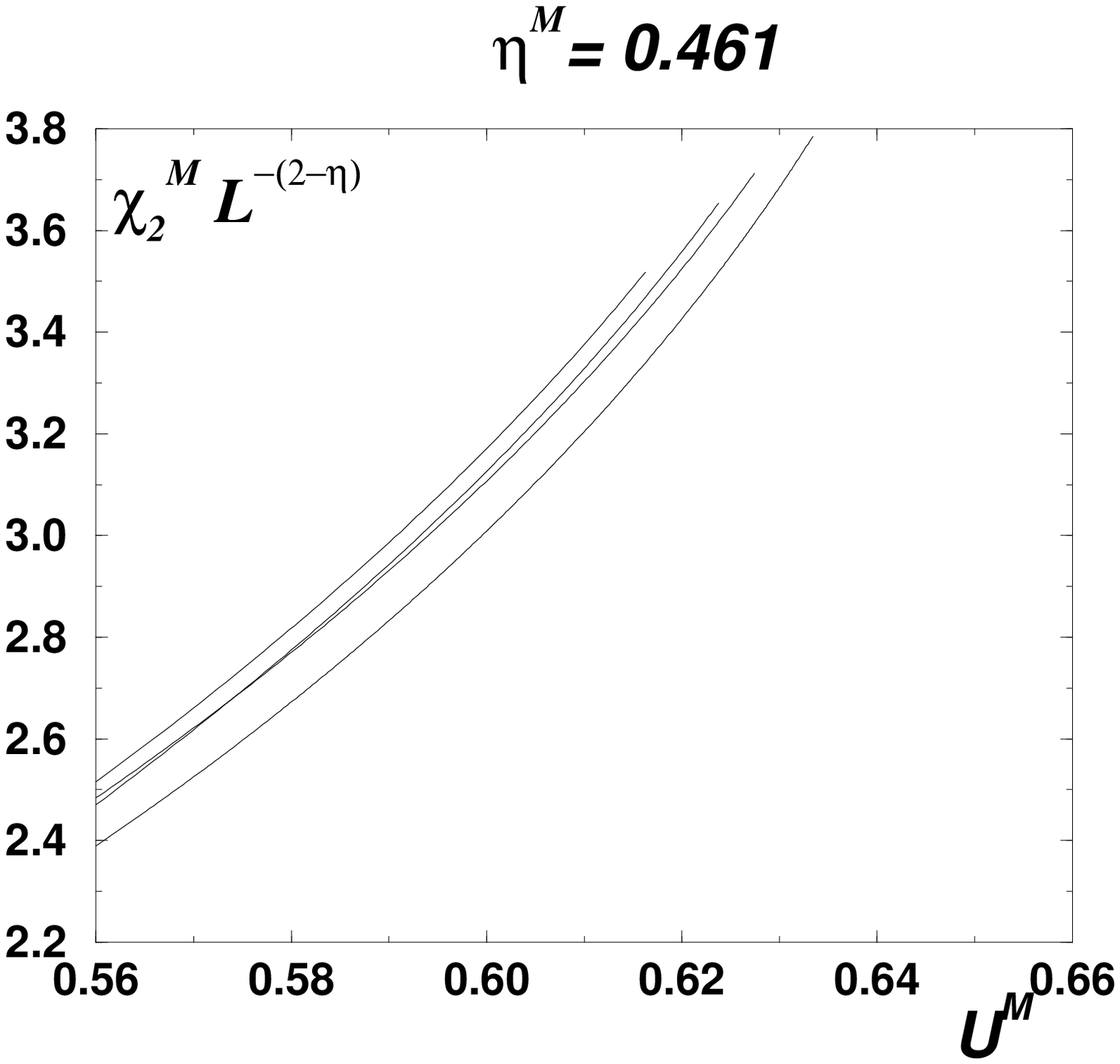,width=6.5cm}
\caption{
\label{X2.5.KT}
}
\vspace{1cm}

\psfig{figure=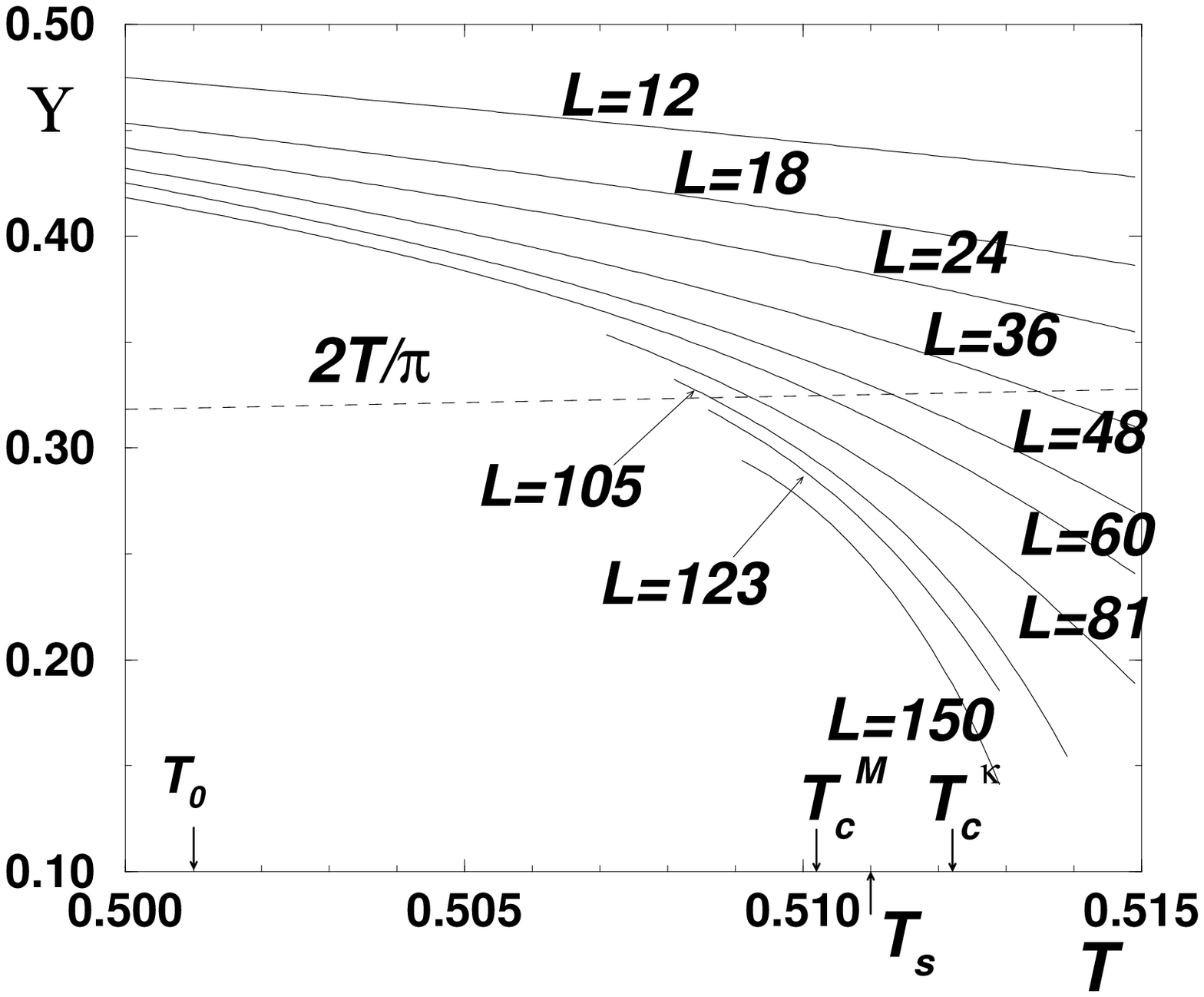,width=6.5cm}
\caption{
\label{fig.HELI}
}
\vspace{1cm}

\psfig{figure=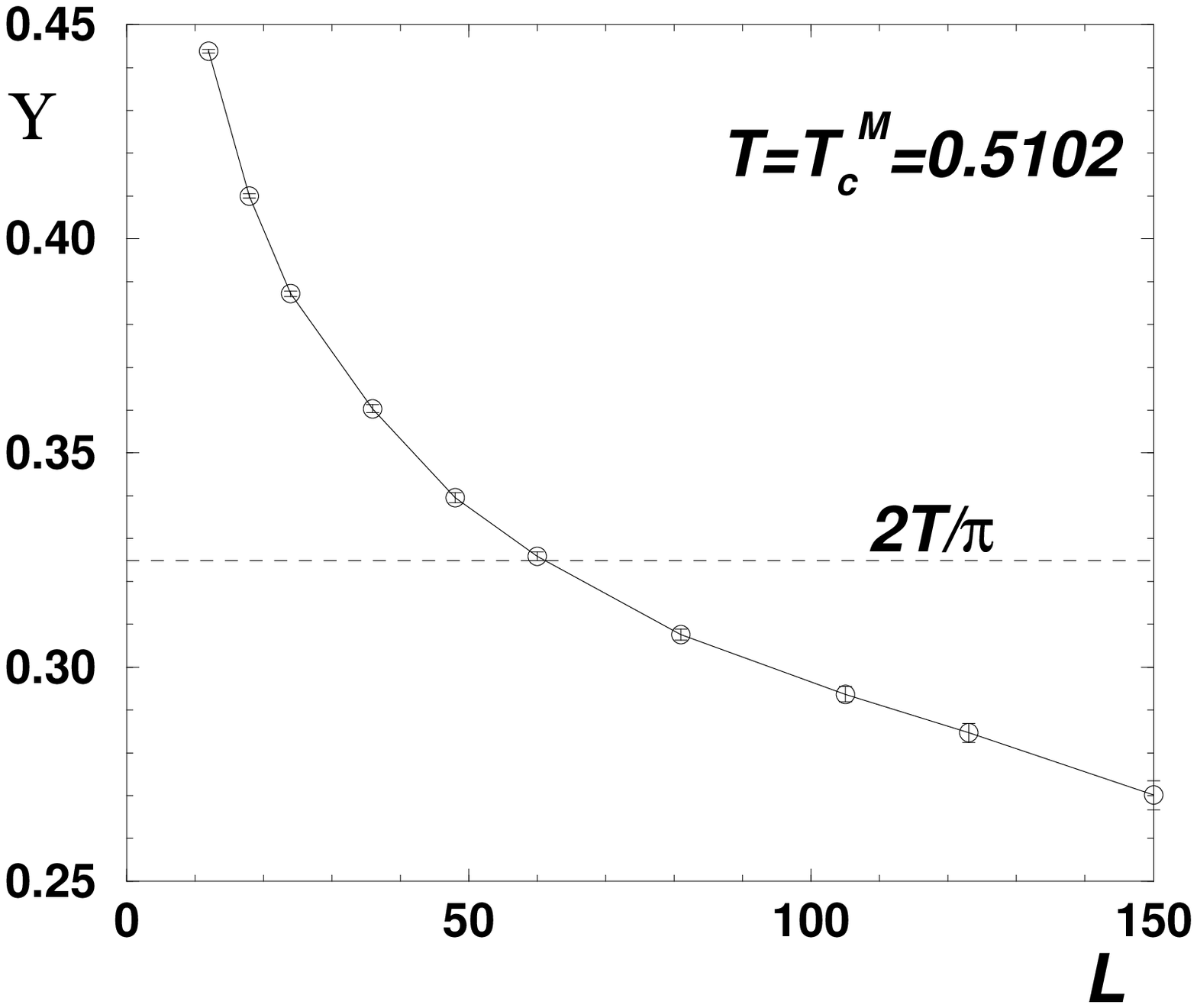,width=6.5cm}
\caption{
\label{fig.HELI.Tc}
}
\vspace{1cm}

\psfig{figure=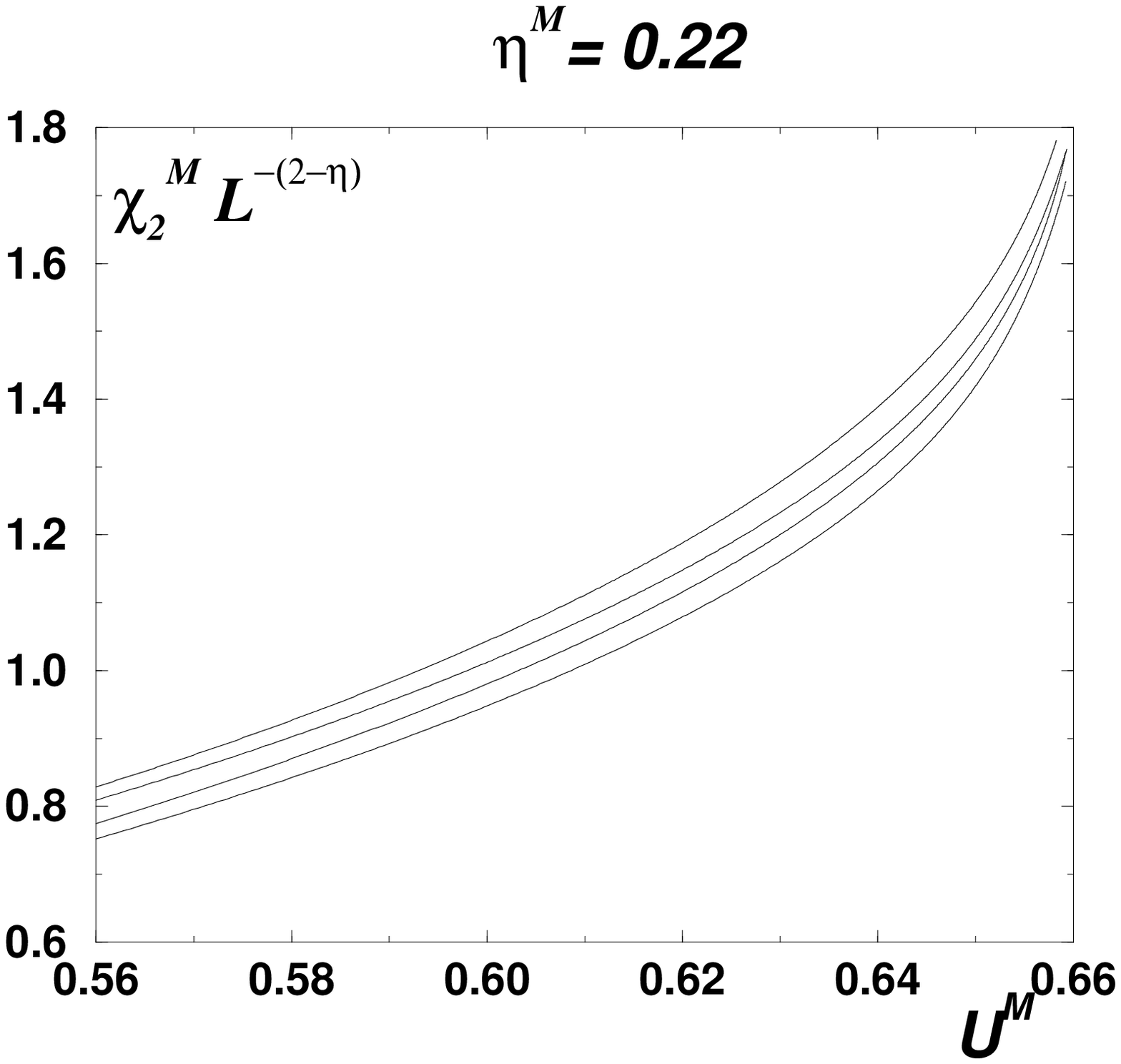,width=6.5cm}
\caption{
\label{X2.4.KT}
}
\vspace{1cm}

\psfig{figure=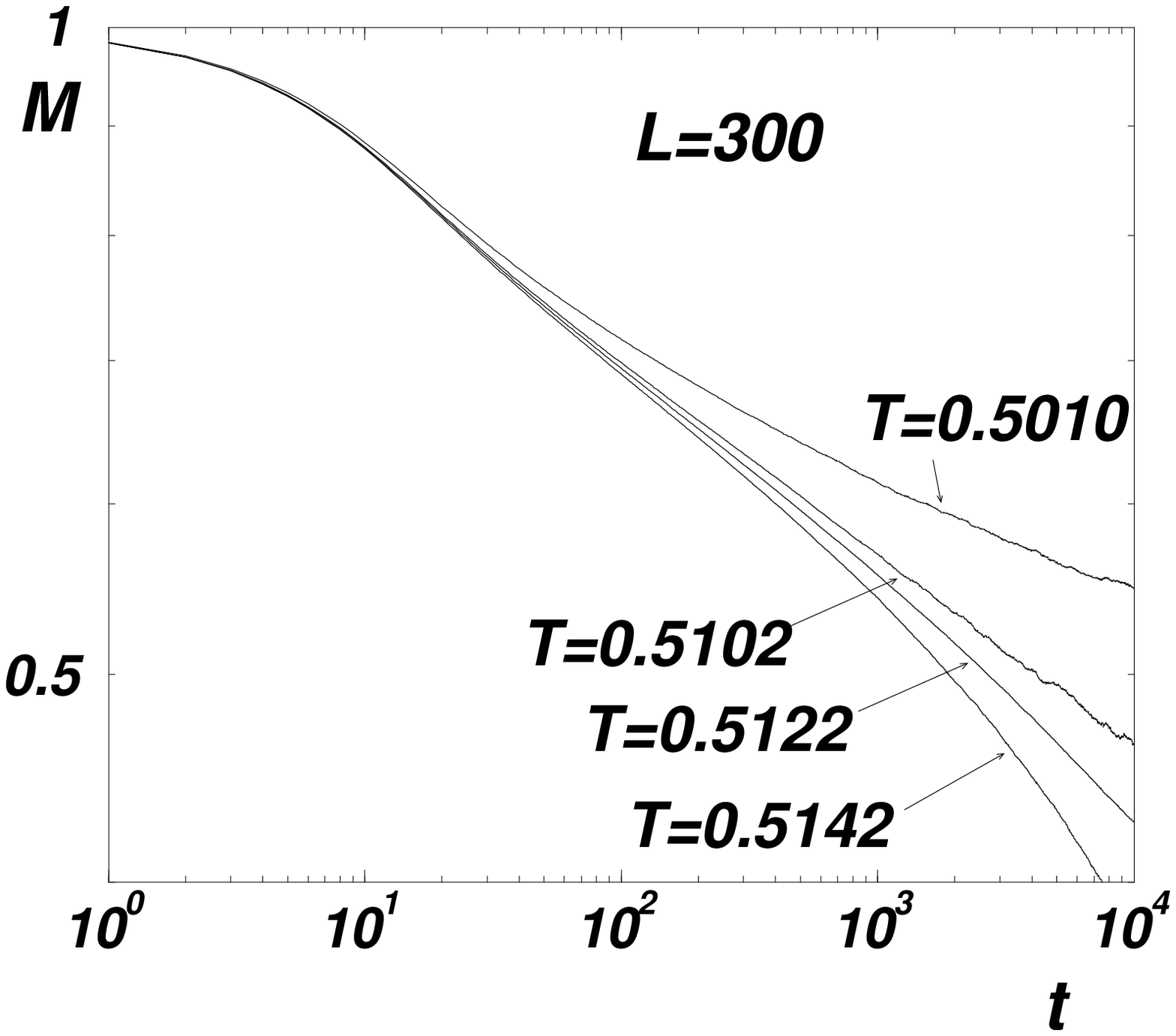,width=6.5cm}
\caption{
\label{fig.a3}
}
\end{center}
\end{figure}

Fig.~\ref{fig.GS}:
Ground state configuration for the triangular model.
The chirality of each triangle is indicated by $+$ or $-$.

Fig.~\ref{fig.NMET.NOR}:
Autocorrelation time $\tau$ times the number of Metropolis steps $N_{MET}$
as function of $N_{OR}/N_{MET}$. $N_{OR}$ is the number of 
over-relaxation steps.

Fig.~\ref{fig.tau.L}:
CPU time (proportional to the autocorrelation time $\tau$) 
for the standard Metropolis algorithm
($A_{0}$-square), in combination with one over-relaxation step
($A_{1}$-diamond),
in combination with $c\,L\sim0.22\,L$ over-relaxation steps 
($A_{cL}$-circle),
and for the Wolff's cluster algorithm ($A_w$-triangle).

Fig.~\ref{figCM}:
Binder's parameter $U^\kappa$ for the Ising order parameter
function of the temperature when $L=12$ to 150.
The arrow shows the temperature of simulation $T_s=0.511$.

Fig.~\ref{figTc2.all}:
Crossing $T$ plotted vs inverse logarithm of the scale factor $b=L'/L$.
The upper part of the figure corresponds to $U^\kappa$ while the lower
part  to $U^M$. 
We obtain $T_c^\kappa=0.5122(1)$ and
$T_{c}^M= 0.5102(1)$ with a linear fit (see text for comments).

Fig.~\ref{figV}:
Values of $V_1^\kappa$ and $V_2^\kappa$ as function of $L$ in log--log
scale at $T_c^\kappa$. The value of the slopes gives $1/\nu^\kappa$.
Strong corrections for  small sizes are visible. Only the three largest sizes
are used for the fits.
The estimated statistical errors are smaller than the symbols.

Fig.~\ref{figX.all}:
Values of $\chi^\kappa$ and $\chi_2^\kappa$  as function of $L$ in log--log
scale at $T_c^\kappa$ and $\chi_2^M$ at $T_c^M$.
The slopes give $\gamma/\nu=2-\eta$. 
Strong corrections for the small sizes for $\chi^\kappa$ are visible. 
Only the three largest sizes are used for the fit for $\chi^\kappa$
and the four largest for  the fits for $\chi_2^\kappa$ and $\chi_2^M$.
The estimated statistical errors are smaller than the symbols.

Fig.~\ref{figM}:
Values of $\langle \kappa\rangle$ as function of $L$ in log--log
scale at $T_c^\kappa$. The slope gives $\beta^\kappa/\nu^\kappa$.
Strong corrections for the small sizes are visible. 
Only the three largest sizes
are used for the fits. The estimated
statistical errors are smaller than the symbol.

Fig.~\ref{X2.1}:
$\chi^\kappa L^{-\gamma^\kappa/\nu^\kappa}$ as function of $U^\kappa$ with
$\gamma^\kappa/\nu^\kappa=1.82$ for sizes from $L=24$ to 150.
The curves do not collapse in one curve.

Fig.~\ref{X2.2}:
$\chi^\kappa L^{-\gamma^\kappa/\nu^\kappa}$ as function of $U^\kappa$ with
$\gamma^\kappa/\nu^\kappa=1.78$ for sizes from $L=24$ to 150.
The curves collapse in one curve.

Fig.~\ref{X2.3}:
$\chi^\kappa L^{-\gamma^\kappa/\nu^\kappa}$  as function of $U^\kappa$ with
$\gamma^\kappa/\nu^\kappa=1.74$ for sizes from $L=24$ to 150.
The curves do not collapse in one curve.

Fig.~\ref{fig.eta}:
$\eta^\kappa$ and $\eta^M$ as function of critical temperature chosen.
$\Delta\eta^\kappa$ and $\Delta\eta^M$ are the estimate
from a direct fit of $\chi$ as function of $U$. This give estimates of the critical
temperatures in agreement with (\ref{eqn.Tc.K}) and (\ref{eqn.Tc.M}). 
Black circles show the result using dynamical properties.

Fig.~\ref{fig.a2}:
$\langle \kappa\rangle$ as function of the time $t$ for various system sizes at the 
critical temperature $T_c^\kappa=0.5122$. The curves for the biggest
sizes $L=201$ and $L=300$ collapse within the errors.

Fig.~\ref{fig.a4}:
$\langle \kappa\rangle$ as function of the time $t$ for temperatures around 
$T_c^\kappa=0.5122$. A linear behavior appears only for $T=T_c^\kappa$.

Fig.~\ref{fig.U.t}: 
Binder cumulant $U^\kappa$ as function of time $t$ at $T_c^\kappa=0.5122$. 
The slope gives $d/z^\kappa$. Only data between $t=300$ and $t=10,000$ 
(shown by the arrows) is used for the fit. 

Fig.~\ref{fig.U.X.V1}: 
Binder cumulant $U^\kappa$ and susceptibility $\chi^\kappa$  as function
of $V_1^\kappa$ at $T_c^\kappa=0.5122$. The slopes  
give $\nu^\kappa$ and $\gamma^\kappa$ respectively. 
We use only data between $t=300$ and $t=10,000$ 
(shown by the arrows) for the fit. 

Fig.~\ref{fig.X.U}: 
$\chi^\kappa$  as function of the binder cumulant $U^\kappa$ 
at $T_c^\kappa=0.5122$. The slope
gives $(2-\eta^\kappa)/d$. 
We use only data between $t=300$ and $t=10,000$ 
(shown by the arrows) for the fit. 

Fig.~\ref{fig.I-XY}:
Phase diagram of the Ising-$XY$ model.
Solid and dotted lines indicate continuous and first-order
transitions respectively. The filled black circles show the possible 
fixed points.
$A$ and $C$ are free  parameter. Similar phase diagram appears for 
the $J_1-J_2$ model\cite{Loison.Simon.J1J2}, 
the Zig-Zag model\cite{Benakly97,Boubcheur98}
and the RSOS model\cite{LeeLeeRSOS}.

Fig.~\ref{figCM.KT}:
Binder's parameter $U^M$ for the $U(1)$ order parameter
function of the temperature for various sizes $L$.
The arrow shows the temperature of simulation $T_s=0.511$.
The scales is similar to those of Fig.~\ref{figCM}

Fig.~\ref{X2.1.KT}:
$\chi^M L^{2-\eta^M}$  as function of $U^M$ with
$\eta^M=0.34$ for the sizes $L=~81,~105,~123$ and 150.
The curves do not collapse in one curve.

Fig.~\ref{X2.2.KT}:
$\chi^M L^{2-\eta^M}$  as function of $U^M$ with
$\eta^M=0.375$ for the sizes $L=~81,~105,~123$ and 150.
The curves collapse in one curve.

Fig.~\ref{X2.3.KT}:
$\chi^M L^{2-\eta^M}$  as function of $U^M$ with
$\eta^M=0.41$ for the sizes $L=~81,~105,~123$ and 150.
The curves do not collapse in one curve.

Fig.~\ref{X2.4}:
$\chi^\kappa L^{-\gamma^\kappa/\nu^\kappa}$  as function of $U^\kappa$ with
$\gamma^\kappa/\nu^\kappa=1.885$ for the sizes from $L=24$ to 150.    
The curves do not collapse in one curve.

Fig.~\ref{X2.5.KT}:
$\chi^M L^{2-\eta^M}$  as function of $U^M$ with
$\eta^M=0.461$ for the sizes $L=~81,~105,~123$ and 150.
The curves do not collapse in one curve.

Fig.~\ref{fig.HELI}:
Helicity $\Upsilon$ for the $U(1)$ symmetry 
as function of the temperature for various sizes $L$.
The arrows show the temperature of simulation $T_s=0.511$,
The critical temperatures for the $U(1)$ symmetry $T_c^M$ 
and the Ising symmetry $T_c^\kappa$, $T_0$ showing the best fit with
the scaling form (\ref{eqn.HELI.L}).
The dashed line $2T/\pi$ represents the universal ferromagnetic jump.

Fig.~\ref{fig.HELI.Tc}:
Helicity $\Upsilon$ for the $U(1)$ symmetry 
as function of the lattice size $L$ at $T_c^M$.
The jump is smaller than the universal ferromagnetic jump (dashed line).

Fig.~\ref{X2.4.KT}:
$\chi^M L^{2-\eta^M}$  as function of $U^M$ with
$\eta^M=0.22$ for the sizes $L=~81,~105,~123$ and 150.
The curves do not collapse in one curve.

Fig.~\ref{fig.a3}:
$\langle M\rangle$ as function of the time $t$ for various temperature.
A linear behavior appears only for $T=T_c^M=0.5102$. 
The deviation from a linear behavior is clearly visible for $T=T_0=0.5010$.

\newpage

\end{document}